\documentclass[prl,superscriptaddress,reprint,showpacs,amsmath,amssymb,nofootinbib%
,nobalancelastpage]{revtex4-1}

\usepackage{xcolor}
\usepackage{graphicx}
\usepackage{subfigure}
\usepackage[colorlinks=true,urlcolor=blue,linkcolor=blue]{hyperref}

\def\({\left(}
\def\){\right)}

\def\ad{a^\dagger}

\def\eq#1{Eq.~(\ref{eq:#1})}

\def\eqs#1#2{Eqs.~(\ref{eq:#1}) \& (\ref{eq:#2})}
\def\eqlist#1#2{Eqs.~(\ref{eq:#1}-\ref{eq:#2})}
\def\Eqs#1#2{Equations~(\ref{eq:#1}) \& (\ref{eq:#2})}

\def\fig#1{Fig.~\ref{fig:#1}}

\def\Fig#1{Figure~\ref{fig:#1}}

\makeatletter
\@ifundefined{textcolor}{}
{%
 \definecolor{BLACK}{gray}{0}
 \definecolor{WHITE}{gray}{1}
 \definecolor{RED}{rgb}{1,0,0}
 \definecolor{GREEN}{rgb}{0,.4,0}
 \definecolor{BLUE}{rgb}{0,0,1}
 \definecolor{CYAN}{cmyk}{1,0,0,0}
 \definecolor{MAGENTA}{cmyk}{0,1,0,0}
 \definecolor{YELLOW}{cmyk}{.2,.4,1,0}
 }
\makeatother

\graphicspath{{../Graphics3/}}

\def\id{I}

\def\1{\mat{\id}}

\def\mat#1{\bm{\mathrm{#1}}}

\begin{document}

\title{Experimental realization of multipartite entanglement of 60 modes of a quantum optical frequency comb}
\author{Moran Chen}
\affiliation{Department of Physics, University of Virginia, Charlottesville, Virginia 22903, USA}
\author{Nicolas C. Menicucci}
\email{ncmenicucci@gmail.com}
\affiliation{School of Physics, The University of Sydney, Sydney, NSW 2006, Australia}
\author{Olivier Pfister}
\email{opfister@virginia.edu}
\affiliation{Department of Physics, University of Virginia, Charlottesville, Virginia 22903, USA}
%\pacs{03.67.-a, 03.67.Pp, 42.50.Dv, 42.50.Ex, 42.50.Lc, 42.65.Yj}

\date{\today}

\begin{abstract} %600 characters only
We report the experimental realization and characterization of one 60-mode copy, and of two 30-mode copies, of a dual-rail quantum-wire cluster state in the quantum optical frequency comb of a bimodally pumped optical parametric oscillator. This is the largest entangled system ever created whose subsystems are all available simultaneously. The entanglement proceeds from the coherent concatenation of a multitude of EPR pairs by a single beam splitter, a procedure which is also a building block for the realization of hypercubic-lattice cluster states for universal quantum computing.
\end{abstract}

\maketitle

%------------------------------------------------------------------------------------------------------------
\textit{Introduction.}---%
Initially identified by Einstein, Podolsky, and Rosen (EPR)~\cite{Einstein1935} as central to testing the completeness of quantum mechanics, entanglement is  also crucial to exponential speedups {of} quantum computing~\cite{Feynman1982,Shor1994,Nielsen2000,Hallgren2007}. In the race to build a practical quantum computer~\cite{Ladd2010}, the ability to create very large quantum registers and entangle them is paramount, along with the ability to address the issue of decoherence. The study of large-scale entanglement---i.e., multipartite entanglement between numerous subsystems---is in itself an intriguing topic at the forefront of current research, as such systems have yet to be studied in laboratories.

Until recently, the largest entangled state of any sort involved 14 trapped ions~\cite{Monz2011}. {Quantum optical systems, which suffer less from decoherence but are harder to entangle, have shown progress, with photon-based, discrete-variable implementations of a 4-qubit ``compiled,'' nonscalable version of Shor's algorithm~\cite{Lanyon2007,Lu2007}, including in an integrated optics platform~\cite{Politi2009}, 4-qubit blind quantum computing~\cite{Barz2012}, and 8-qubit topological quantum error correction~\cite{Yao2012}.

With particular regard to scalability, the field-based, continuous-variable (CV) flavor of quantum optics has high potential~\cite{Pfister2004,Menicucci2007,Menicucci2008,Flammia2009,Patera2012}, in particular by enabling ``top down,'' rather than ``bottom up,'' entangling approaches of quantum field modes. It is also important to note the relevance of continuous variables to universal quantum computing, with the recent discovery of a fault tolerance threshold for quantum computing with CV cluster states and nonGaussian error correction~\cite{Menicucci2013a}.

In 2011, 15 independent 4-mode cluster states were generated simultaneously over 60 modes of the quantum optical frequency comb (QOFC) of a single optical parametric oscillator (OPO)~\cite{Pysher2011}. {In 2013}, 10-mode entanglement was {observed} in a synchronously pumped OPO~\cite{Roslund2014}, and 10,000 modes were sequentially entangled into a dual-rail cluster state~\cite{Yokoyama2013} following a time-domain protocol~\cite{Menicucci2010,Menicucci2011a} in which the modes are emitted in pairs and detected in turn, with only a few modes accessible at any given time.

In this Letter, we report the experimental multipartite entanglement of 60 adjacent modes of the QOFC of a single OPO, all simultaneously available. The number of entangled modes was limited by our measurement technique, not by the generation process (which, we estimate~\cite{Wang2013a}, yielded in excess of 6,000 entangled modes). This is the largest entangled state ever created in which all constituent systems are simultaneously available and addressable. Moreover, the entanglement is not of an arbitrary type (e.g., largely due to experimental convenience~\cite{Pysher2011,Roslund2014}) but a carefully engineered, sophisticated resource---a continuous-variable dual-rail quantum wire~\cite{Alexander2013}---that has direct applications in quantum computing~\cite{Menicucci2006,Wang2013} and in experimental studies of topological order in quantum many-body systems~\cite{DeMarie2013}---a novel quantum phenomenon that hasn't yet been revealed experimentally. Beyond these immediate applications, it also forms a basic building block for much larger entangled states with rich, regular-lattice structure~\cite{Wang2013}, some of which could not otherwise be embedded in three-dimensional space. The intrinsic scalability of the experimental design paves the way for a new program of experimental research into the properties and applications of these richly entangled multipartite quantum systems.

%------------------------------------------------------------------------------------------------------------
\textit{Principle of the experiment.}---%
The QOFC was formed by the resonant modes of the optical cavity of a {doubly resonant} OPO. The OPO contained periodically poled $\rm KTiOPO_{4}$ (PPKTP) nonlinear crystals which quasiphasematched $zzz$ parametric downconversion (PDC)---the concurrent annihilation of a $z$-polarized pump photon at the 532 nm wavelength and  creation of a $z$-polarized photon pair at the 1064 nm wavelength. Due to the cavity's resonant enhancement, the signal pair frequencies,  add{ing} up to the pump frequency, {a}re the cavity eigenfrequencies{, at which higher-photon-flux} PDC {yields} two-mode squeezing,  the bipartite entanglement mechanism of EPR pairs~\cite{Ou1992}. {Our} OPO was polarization-degenerate: its 2 identical, $x$-cut PPKTP crystals were oriented at $90^{\circ}$ from each other in the ($yz$) plane, leading to the generation of two distinct sets of EPR pairs, $zzz$ and $yyy$, as depicted in \fig{OneBelt}.
\begin{figure}[t!]
\centering
\subfigure{
\includegraphics[width=0.42\textwidth]{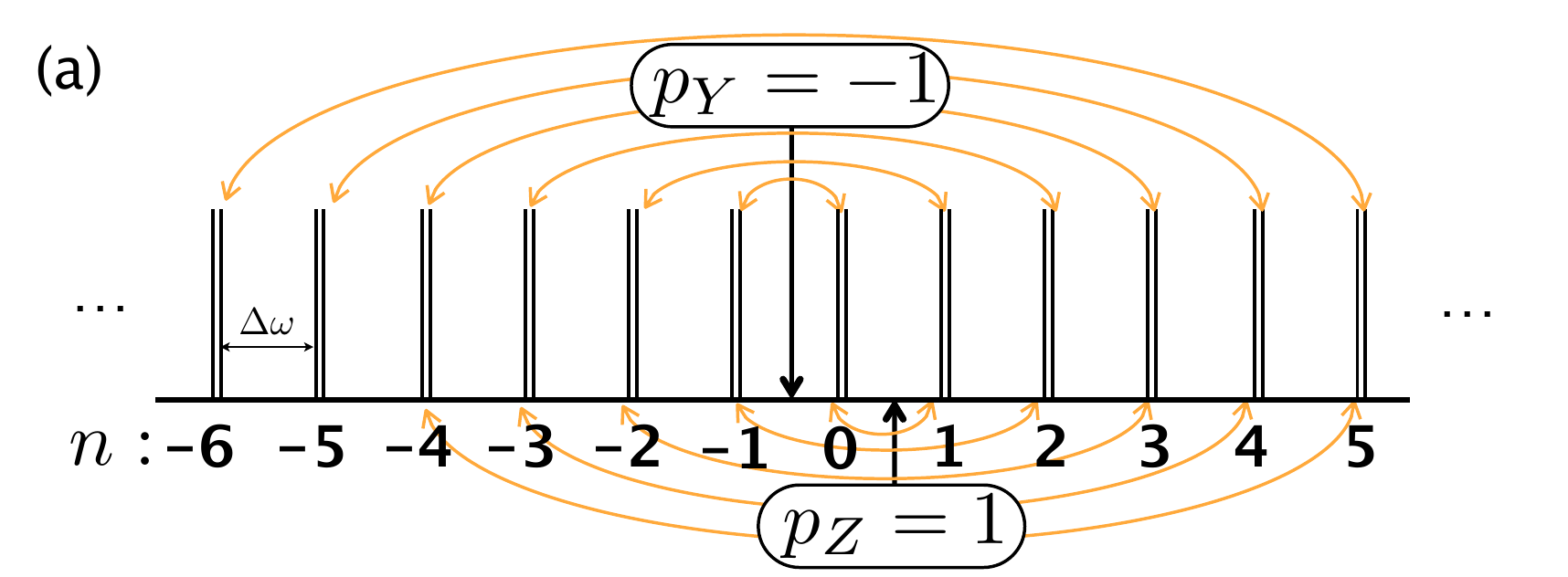}
  \label{fig:OneBelt}
}
\subfigure{
\includegraphics[width=0.39\textwidth]{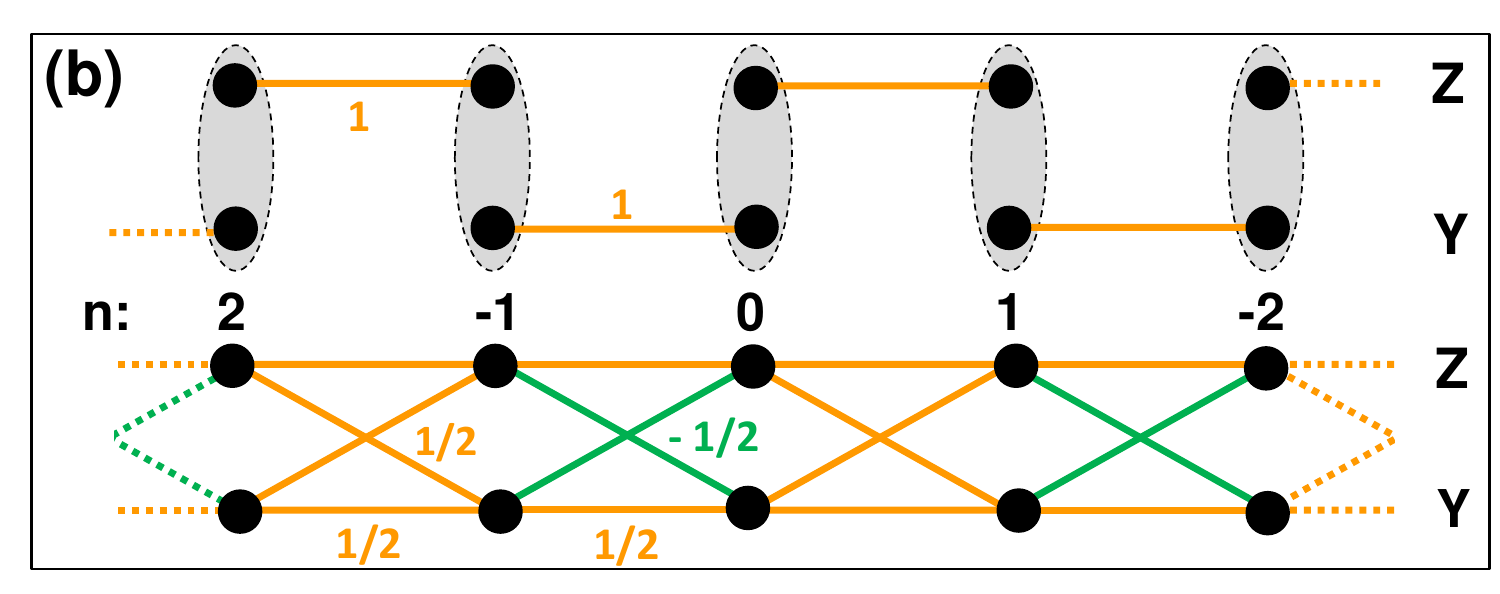}
  \label{fig:EntBelt}
}
\caption{Generation of a dual-rail quantum wire in the QOFC. (a) EPR pairs created by $zzz$ and $yyy$ interactions in the QOFC of a polarization-degenerate OPO (at each frequency $n$ the $z$ and $y$ modes are denoted by the double black lines). The vertical arrows mark the half-frequencies of the pumps; the curved arrows denote the $zzz$ (bottom) and $yyy$ (top) EPR pairs. (b) Quantum graph states~\cite{Menicucci2011}: The initial EPR pairs from the OPO (top) turn, after a single beam splitter (grey ellipses), into a dual-rail CV cluster state (bottom), whose $\pm1/2$-weight edges  are color-coded (contrary to the qubit case, weighted cluster CV states are still stabilizer states~\cite{Gu2009,Menicucci2011}).}
%\vglue -0.1in
\label{fig:Belt}
\end{figure}
We label the modes in the QOFC by integer $n$ such that $\omega_{n}=\omega_0 + n \Delta\omega$, with $\omega_0$ an arbitrary origin and $\Delta\omega$ the OPO free spectral range (FSR). The PDC phasematching condition for EPR pair ($n_{1},n_{2}$) gives $\omega_{p}=\omega_{n_1} + \omega_{n_2} = 2\omega_0 + p\Delta\omega$, where $p=n_{1}+n_{2}$ is the pump index. For $|p_{y}-p_{z}|=2$, i.e., pump frequencies differing by exactly twice the OPO FSR, {all} EPR pairs concatenate into the frequency sequence ($\ldots,-6,5,-4,3,-2,1,0,-1,2,-3,4,-5,\ldots$) [\fig{OneBelt}] that extends to the whole phasematching bandwidth in the QOFC. We recently measured the latter to be {more than} 3.2 THz-wide~\cite{Wang2013a}. Hence, we estimate that our entangled QOFC, of mode spacing $\Delta\omega=0.95$ GHz, extends over at least $2N=6,700$ modes (counting both polarizations).

This frequency sequence yields frequency-staggered EPR pairs in \fig{EntBelt}, top. As was  shown {for}  sequential CV entanglement~\cite{Menicucci2011a,Yokoyama2013}, a balanced beam splitter {entangles} EPR pairs (which are also CV cluster states, {up to local phase shifts}), temporally staggered by an optical delay line, into the dual-rail CV cluster state depicted in \fig{EntBelt}, bottom. In our work, the staggering of the EPR pairs is spectral, caused by the decoherence-free pump frequency splitting.

To verify entanglement, we measured the joint squeezed operators called \textit{variance-based entanglement witnesses}~\cite{Hyllus2006} and \textit{nullifiers}~\cite{Gu2009}, which are the solutions of our OPO's Heisenberg equations. Nullifiers are directly related to the stabilizers of the generated cluster state~\cite{SuppMat} and are also used in a more general entanglement check by the van Loock-Furusawa criterion~\cite{vanLoock2003a}.
  {Their derivation} in the Heisenberg picture (see also Refs.~\onlinecite{Menicucci2011a,Yokoyama2013,Wang2013}){uses t}he OPO's interaction-picture Hamiltonian{,}
\begin{align}
H =  i\hbar\left[\kappa_{z}\sum_{k=n_{z}}^{\frac N2}a^{(z)\dagger}_{k}a^{(z)\dagger}_{p_{z}-k}
 + \kappa_{y}\sum_{l=n_{y}}^{\frac N2}a^{(y)\dagger}_{l}a^{(y)\dagger}_{p_{y}-l}\right]+ H.c.,
\end{align}
where $n_{z,y}=\lceil \frac{p_{z,y}}2 \rceil$, {whose} well-known solutions  are the EPR nullifiers $[Q^{(j)}_{n}-Q^{(j)}_{p_{j}-n}]e^{-r_{j}}$ and $[P^{(j)}_{n}+P^{(j)}_{p_{j}-n}]e^{-r_{j}}$, $j=y,z$, where $r_{j}=\kappa_{j}t$ are the squeezing parameters. {A} $45^{\circ}$ polarization rotation matrix $\left( \begin{smallmatrix} 1 & 1 \\ 1 & -1 \end{smallmatrix} \right)/\sqrt2$, applied to annihilation operators $(a_{n}^{(z)}, a_{n}^{(y)})^{T}$, transforms the EPR nullifiers into
\begin{align}
Q_{p_{z}-n,n}(r_{z})&=\{[Q^{(z)}_{n}+Q^{(y)}_{n}]-[Q^{(z)}_{p_{z}-n}+Q^{(y)}_{p_{z}-n}]\}e^{-r_{z}}\label{eq:nbs1}\\
P_{p_{z}-n,n}(r_{z})&=\{[P^{(z)}_{n}+P^{(y)}_{n}]+[P^{(z)}_{p_{z}-n}+P^{(y)}_{p_{z}-n}]\}e^{-r_{z}}\label{eq:nbs2}\\
Q_{p_{y}-n,n}(r_{y})&=\{[Q^{(z)}_{p_{y}-n}-Q^{(y)}_{p_{y}-n}]-[Q^{(z)}_{n}-Q^{(y)}_{n}]\}e^{-r_{y}}\label{eq:nbs3}\\
P_{p_{y}-n,n}(r_{y})&=\{[P^{(z)}_{p_{y}-n}-P^{(y)}_{p_{y}-n}]+[P^{(z)}_{n}-P^{(y)}_{n}]\}e^{-r_{y}}.\label{eq:nbs4}
\end{align}
Assuming\footnote{See supplemental material~\cite{SuppMat} for an analysis of deviations from this case.} $r_{z}=r_{y}=r$, taking the sum and difference of \eqs{nbs1}{nbs3} and applying a Fourier transform---a.k.a.\ a local $\frac\pi2$ optical phase shift---to mode $n$ yields the {standard} CV graph nullifiers\footnote{\Eqs{nbs2}{nbs4} are unused for graph node $n$ and for all others of the same parity ($n\pm2$...). They are the sole starting point for the nullifier derivations for graph nodes of opposite parity ($n\pm1$...)}
\begin{align}
\left[P^{(z)}_{n} - \frac12 (Q^{(y)}_{p_{z}-n}+Q^{(z)}_{p_{z}-n}+Q^{(z)}_{p_{y}-n}-Q^{(y)}_{p_{y}-n})\right]e^{-r}\label{eq:nps1f}\\
\left[P^{(y)}_{n} - \frac12 (Q^{(y)}_{p_{z}-n}+Q^{(z)}_{p_{z}-n}-Q^{(z)}_{p_{y}-n}+Q^{(y)}_{p_{y}-n})\right]e^{-r}\label{eq:nps2f}
\end{align}
which correspond exactly to \fig{EntBelt}, bottom~\cite{Gu2009,Menicucci2011}. The measurement of these nullifiers requires homodyne detection at 3 different optical frequencies. However, one may also measure the more convenient observables of \eqlist{nbs1}{nbs4}, displayed in \fig{colorchain}, which only require the two-tone homodyne detection implemented in Ref.~\onlinecite{Pysher2011}.
\begin{figure}[t!]
\includegraphics[width=0.2\textwidth]{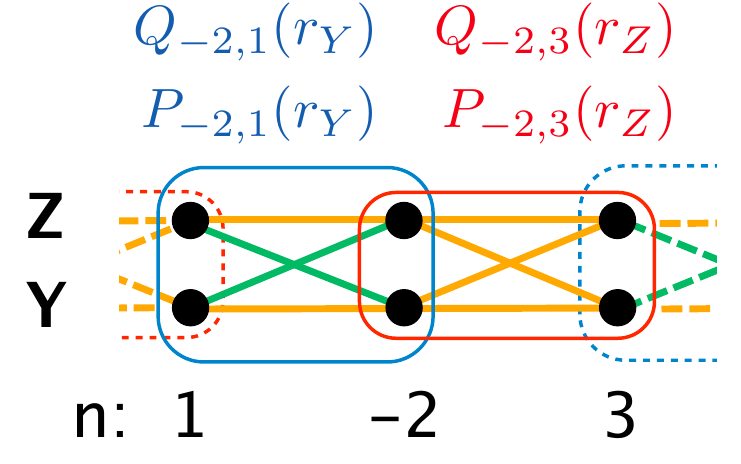}
\caption{Visualization of the measured nullifiers of \eqlist{nbs1}{nbs4} (blue and red boxes) on the dual-rail graph state of \fig{EntBelt}. As shown in the text, simultaneous squeezing of $Q_{-2,3}(r_z)$ and $Q_{-2,1}(r_y)$ is equivalent to squeezing of the canonical nullifiers of \eqs{nps1f}{nps2f}.}
\label{fig:colorchain}
\end{figure}

A remarkable feature of our frequency-domain implementation is that merely tuning the pump spacing{,} $|p_{y}-p_{z}|=2m$, {yields} $m$ disjoint frequency sequences and hence $m$ independent dual-rail cluster states. See \fig{TwoBelts} for  $m=2$, {implemented} in this work along with $m=1$ {(}\fig{Belt}{)}. Note that {all} nullifier measurements are  two-tone in {both cases, a} simplification of the experimental procedure {which} is central to our proposed generalization of this work to the generation of CV cluster states with hypercubic lattices~\cite{Wang2013}.

%------------------------------------------------------------------------------------------------------------
\textit{Experimental setup.}---%
Our polarization-degenerate OPO had a bowtie cavity (\fig{setup}) of FSR $\Delta\omega=945.66$ MHz. The OPO cavity length was actively stabilized by locking to a weak counterpropagating beam via a Pound-Drever-Hall (PDH) servo loop. The cavity eigenmode had two waists, where we placed the two PPKTP crystals, one {(31 $\mu$m)} between the curved mirrors and one {(131 $\mu$m)} between the flat mirrors. Great care was taken to suppress polarization crosstalk between the crystals as well as resonant retroreflection from the OPO cavity \cite{SuppMat}.

\begin{figure}[t!]
\centering
\includegraphics[width=\columnwidth]{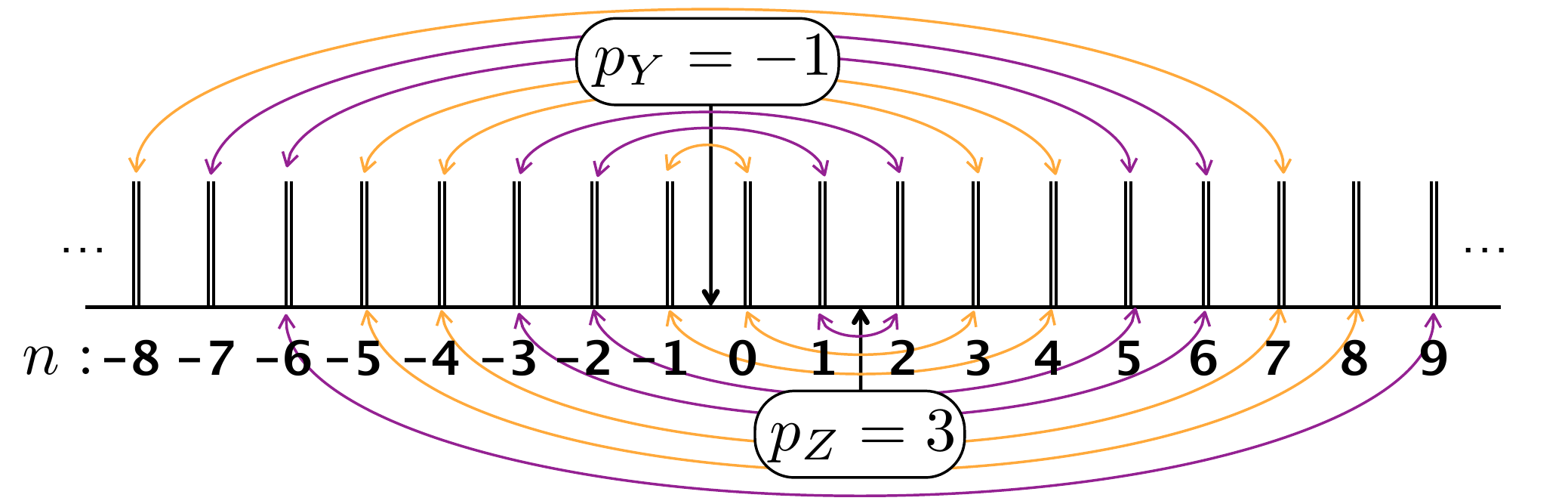}
\caption{Generation of two dual-rail quantum wires. The only difference with \fig{OneBelt} is that the pump frequency difference is $4\Delta\omega$ instead of $2\Delta\omega$. The frequency sequences of the wires are totally distinct: ($\ldots,-8,7,-4,3,0,-1,4,-5,8,\ldots$) for the orange wire and ($\ldots,-7,6,-3,2,1,-2,5,-6,9,\ldots$) for the purple wire.}
\vglue -0.1in
  \label{fig:TwoBelts}
\end{figure}
\begin{figure}[b!]
\centerline{\includegraphics[width=\columnwidth]{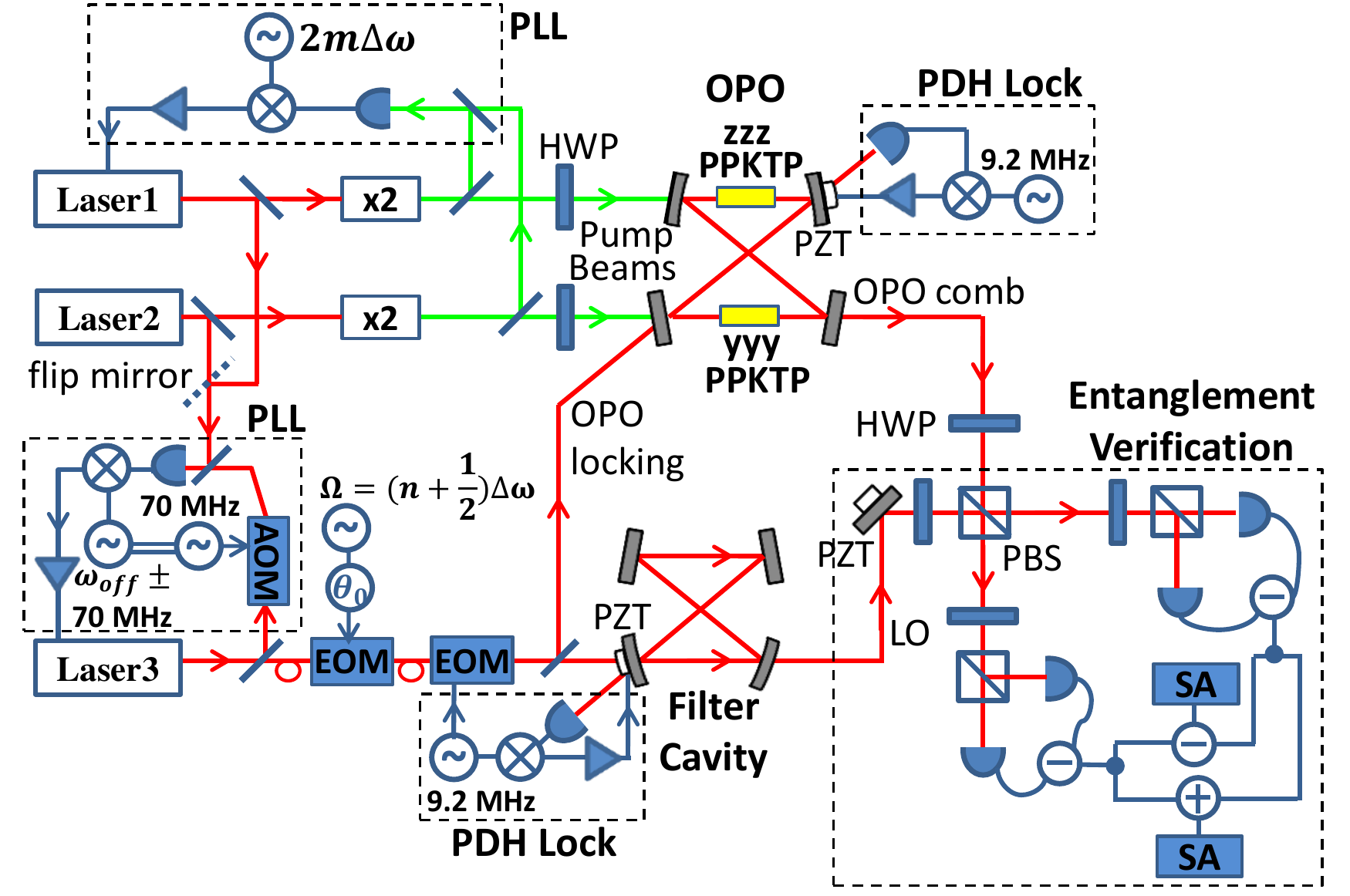}}
\vglue -0.1in
\caption{Experimental setup. PLL: phase-lock loop; HWP: half wave plate; PZT: piezoelectric transducer; PBS: polarizing beam splitter; SA: spectrum analyzer; AOM: acousto-optic modulator; EOM: electro-optic modulator; PDH: Pound-Drever-Hall lock loop.}
\vglue -0.1in
\label{fig:setup}
\end{figure}

Two frequency-doubled, ultrastable continuous-wave (cw) Nd:YAG lasers, of frequency linewidth 1 kHz at 532 nm, were used for the pump fields. The lasers were phaselocked together at a frequency difference $2m\Delta\omega$, with $m=1$ or $m=2$. The two pump beams entered the OPO through different paths to {make a single pass through} the $yyy$ and $zzz$ PPKTP crystals separately. To realize $r_{y}=r_{z}$, the pump powers were independently adjusted to compensate for the different waists at each crystal.

To test the dual-rail wire structure, the 4-mode nullifiers were measured, at all frequencies, by a two-tone balanced homodyne detection (BHD) system whose local oscillator (LO) was provided by another Nd:YAG cw laser, phaselocked at (and sometimes offset from) the half frequency of one of the pumps. The two LO tones were then generated by a phase electro-optic modulator (EOM) at a frequency $\Omega=(n+\frac12)\Delta\omega$, such that $\omega_{LO}+\Omega=\omega_{n}$ and $\omega_{LO}-\Omega=\omega_{p_{y}-n}$, for example. The EOM's $\Omega_\mathrm{max}=14$ GHz bandwidth yielded $n_\mathrm{max}=14$, i.e., $2\times15$ measurable modes (starting from $n=0$) for each polarization. (Replacing this EOM system with two phaselocked, widely tunable 1064 nm diode lasers will give us access to the aforementioned 6,700 modes instead of the current 60.) The first-order EOM sidebands were subsequently bandpass-filtered by a cavity of FSR $\Delta\omega$,  PDH-locked on the LO laser. The LO phase was adjusted by a piezoelectric transducer  mirror and an electronic splitter/combiner network was used to form the nullifier signals.

%------------------------------------------------------------------------------------------------------------
\textit{Results.}---%
We conducted three types of experimental tests: {\em (i)} measurements of the squeezed nullifiers of \eqlist{nbs1}{nbs4}, {\em (ii)} tests of the van Loock-Furusawa CV multipartite entanglement criterion~\cite{vanLoock2003a}, {\em (iii)} tests of non-nullifying observables. The supplemental material~\cite{SuppMat} contains the entire data for all 60 measured modes for $m=1,2$. We present here a qualitative summary of the results. For {\em (i)}, the LO was phaselocked exactly at half the frequency of the $y$ pump to measure $Q_{ij}(r_{y})$,  $P_{ij}(r_{y})$, and likewise for $z$. In two-mode BHD, both the LO phase mirror and the phase $\theta_{o}$ of the EOM drive (\fig{setup}) contribute to determining the measured observable{~\cite{SuppMat}}. In practice, switching between \eqs{nbs1}{nbs2} [and between \eqs{nbs3}{nbs4}] was done by tuning $\theta_{o}$ by $\pm\pi/2$ by simply changing the length of a coaxial cable, yielding identical squeezing signals.

\begin{figure}[t!]
\centering
 \includegraphics[width=\columnwidth]{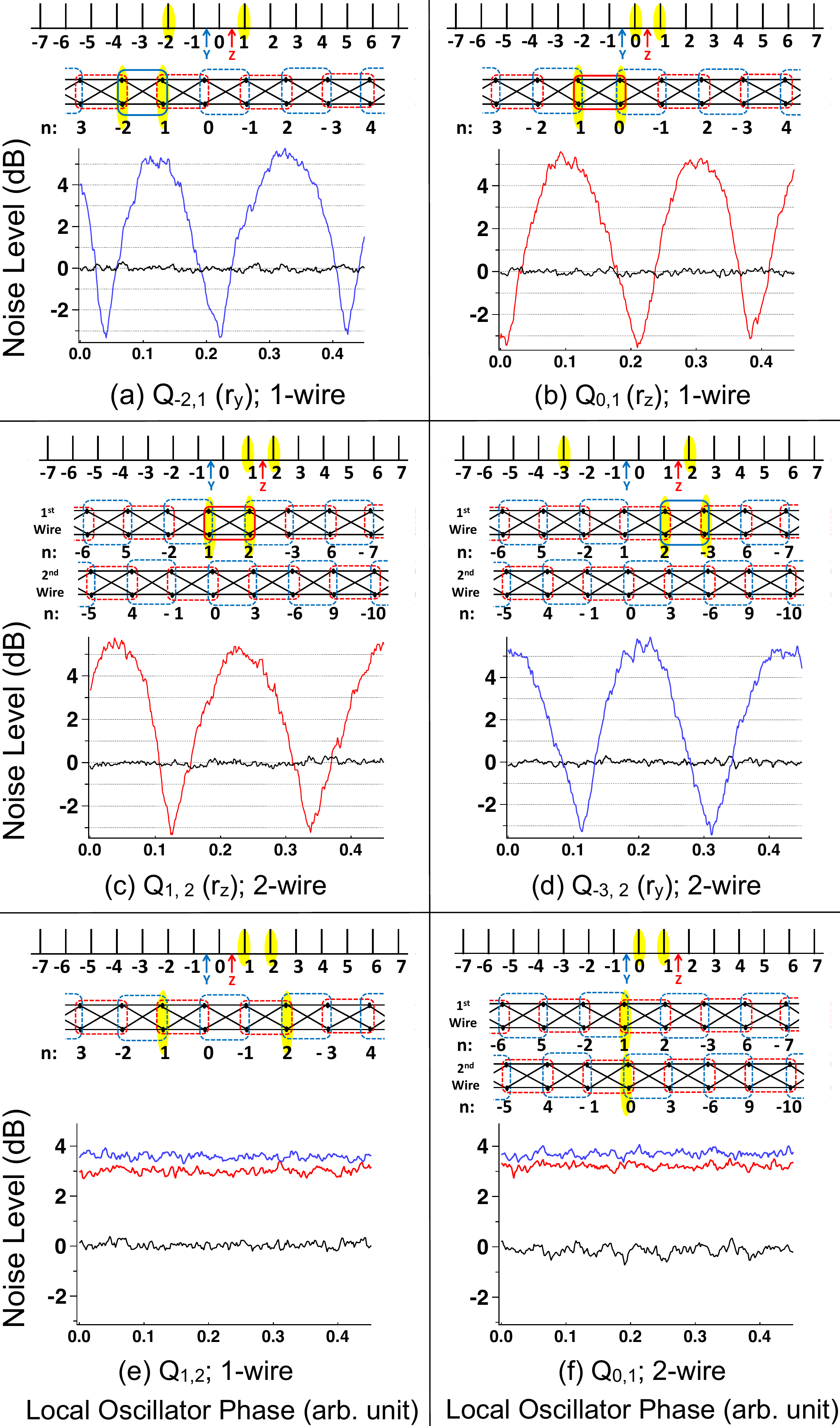}
\caption{Zero-span spectrum analyzer traces of raw squeezing measurements for $m=1$ and $m=2$ quantum wires. For each case, the QOFC is at the top, with the pump half-frequencies denoted by the blue and red arrows, and the quantum graph is beneath it. The yellow highlights indicate the LO sidebands. The black traces indicate the vacuum noise level. Center frequency: 1.25 MHz. Resolution bandwidth: 30 kHz. Video bandwidth: 30 Hz.}
\label{fig:sq}
\end{figure}
\Fig{sq} displays typical squeezing signals in several crucial cases that evidence the graph structure. First, Figs.~\ref{fig:sq}(a,b) prove a ``unit cell'' of the graph, i.e., {which} verifies \eqs{nps1f}{nps2f} for $n=1$  {and} $m=1$. The uncorrected squeezing level was $-3.2\pm 0.2$ dB throughout our measurements. Deconvolving the ``dark'' electronic noise floor of -96 dBm, -13 dB from the vacuum noise level (the LO power was 2 mW at each photodiode), yielded an actual squeezing level of $-3.4\pm 0.2$ dB~\cite{SuppMat}, enough to satisfy the van Loock-Furusawa inseparability criterion level of $-3$ dB~\cite{SuppMat}. The last step {\em (iii)} was to check incorrect graph nodes, exemplified by \fig{sq}(e). The LO was phaselocked at an offset from half the frequency of one pump, which allowed us to measure nullifier observables over the ``wrong'' modes. Phase-independent excess quantum noise was observed, in good agreement with theoretical predictions~\cite{SuppMat}, proving that the measured observable is not a nullifier in this case. The complete set of such {checks} is prohibitively large but all of the ones we tested gave the predicted negative result. All of these measurements demonstrate that a 60-mode dual-rail cluster state was generated in the QOFC.

As {predicted} above, changing the pump splitting to $m=2$ {should} yield two identical {wires}. Figures \ref{fig:sq}(c,d) show measurements demonstrating the ``unit cell'' of one of the wires. Note, in particular, that the successful nullifier measurement of \fig{sq}(c) is the same as that of \fig{sq}(e), which wasn't a nullifier for the $m=1$ pump splitting. Another {such} ``devil's advocate'' check is displayed in \fig{sq}(f), in which cross correlations between the two wires are shown to be absent, even though this very same measurement yielded squeezing for $m=1$ [\fig{sq}(b)]. We confirmed that 2 identical copies of a 30-mode dual-rail cluster state were generated in the QOFC.

%------------------------------------------------------------------------------------------------------------
\textit{Conclusion.}---%
We  demonstrated the ultracompact generation, in a single optical parametric oscillator, of record-size cluster entanglement, thereby realizing the scalability potential of continuous variables in the quantum optical frequency comb. The number of verified entangled modes was limited to 60 by our EOM sideband generation bandwidth. Based on the exceptional  $zzz$ phasematching bandwidth measured in PPKTP at the particular wavelength of 1064 nm~\cite{Wang2013a}, we have strong reason to believe that the maximum number of entangled modes in our experiment is at least 6,700. The OPO is pumped by only two frequencies, in contrast to the complicated spectrum required in our previous proposals~\cite{Menicucci2008,Flammia2009}. In addition, simply tuning the pump frequency difference provides a decoherence-free method for creating multiple independent copies of the same state. The squeezing levels for the one-wire case and two-wire case were identical, showing that the number of copies does not affect their quality. Based on the 60 GHz  emission range of a typical frequency-doubled Nd:YAG pump laser, one can estimate that $m=30$ wires, of $N/30$ modes each, can be created in a  1 GHz-FSR OPO. (Note that using amplified semiconductor lasers as pumps could significantly increase these figures.) We have also shown that interfering several OPOs identical to the one featured in this work should allow one to generate cluster states with hypercubic lattices~\cite{Wang2013}. Finally, another interesting feature of the multiple-copy generation is the availability of states whose entangled modes are widely frequency-spaced (up to 30 GHz in the above estimation), making them accessible for quantum information processing without requiring very high resolution dispersers  \cite{Diddams2007}.

\textit{Acknowldgments.}---%
We thank Pei Wang, Niranjan Sridhar, Matthew Pysher, and Wenjiang Fan for stimulating discussions. This work was supported by the U.S. National Science Foundation under grants No.\ PHY-1206029 and No.\ PHY-0855632. N.C.M. was supported by the Australian Research Council under grant No.~DE120102204.

%------------------------------------------------------------------------------------------------------------
%\bibliography{../Pfister}
%\bibliographystyle{apsrev4-1.bst}

%

\newpage

\onecolumngrid

\part{Appendix: Supplemental material}

\section{Relation between continuous-variable and qubit cluster states}

\subsection{Stabilizers and nullifiers}

The first formulation of the correspondence between continuous-variable and qubit cluster states was given by Zhang and Braunstein~\cite{Zhang2006}. Subsequently, the use of continuous variables for quantum computing, initially proposed by Lloyd and Braunstein~\cite{Lloyd1999}, was furthered for one-way quantum computing by Menicucci et al.~\cite{Menicucci2006s}. Note that first ever determination of a fault tolerance threshold for continuous-variable  measurement-based quantum computing was recently made by Menicucci~\cite{Menicucci2013as}.

Quantum tomography of $N\gg1$ qubits scales exponentially  with $N$ in general, even though efficient techniques have been discovered in a wide range of particular cases~\cite{Cramer2010,Flammia2011,daSilva2011}. For a pure entangled state, an alternative is to measure the $N$ generators of the graph stabilizer group,
\begin{equation}
X_j\bigotimes_{k\in\mathcal N_j}Z_k,
\end{equation}
where $j,k$ denote graph vertices and $\mathcal N_j$ the nearest neighborhood of $j$. By definition, all such operators must have eigenvalue 1 if the quantum state of the system corresponds to the stabilized quantum graph.

When generalizing to CVs, the Hermitian unitary Pauli group generated by $Z$ and $X$ is replaced with the unitary Weyl-Heisenberg group~\cite{Bartlett2002} generated by
\begin{align}
Z(\varpi)&=\exp(i\varpi Q)\\
X(\xi)&=\exp(-i\xi P),
\end{align}
where $P=i(a^{\dagger}-a)/\sqrt2$ and $Q=(a+a^{\dagger})/\sqrt2$ are the phase and amplitude quadratures, respectively. The CV-graph stabilizers are then of the form
\begin{equation}
\exp\left[i\xi\biggl(P_j-\sum_{k\in\mathcal N_j}V_{jk}Q_k\biggr)\right],
\end{equation}
where $V$ denotes the graph's adjacency matrix. The nilpotent Hermitian operator in parentheses is called a {nullifier} \cite{Gu2009s},\footnote{Nilpotency implies infinite squeezing but finite squeezing still constitutes an unequivocal signature of entanglement~\cite{Menicucci2011s,Menicucci2013as}.} or a {variance-based entanglement witness} \cite{Hyllus2006s}, and can be shown to coincide exactly, up to local phase shifts, to the solutions of the Heisenberg evolution equations of the OPO we used.

Moreover, the nullifiers can also be used to implement the more general van Loock-Furusawa inseparability criterion~\cite{vanLoock2003as}.

\subsection{The van Loock-Furusawa inseparability criterion}\label{LF}
This entanglement criterion is the generalization to the multipartite case of the Duan-Simon criterion~\cite{Duan2000,Simon2000}, which is itself the CV version of the Peres-Horodecki criterion~\cite{Peres1996,Horodecki1996}.

We use the van Loock-Furusawa (vLF) separability inequalities~\cite{vanLoock2003as,Yokoyama2013s}. We consider all possible separable bipartitions in our set of entangled modes and  enumerate the necessary conditions for the separability. If the inequalities for the necessary conditions of separability for all the cases are violated, we obtain the sufficient conditions for the full inseparability.

A key point here is that we are dealing with cluster states, in which quantum correlations only involve the nearest neighbors. Thus, we only need to examine the separability of the latter and may use the graph nullifiers as the test observables in building the vLF inequalities.

As was already detailed in the supplemental material of Ref.~\cite{Pysher2011s} (see also Refs.\cite{Leuchs2009,Bloomer2010}), the nullifiers
\begin{align}
Q_{p_{z}-n,n}(r_{z})&=\{[Q^{(z)}_{n}+Q^{(y)}_{n}]-[Q^{(z)}_{p_{z}-n}+Q^{(y)}_{p_{z}-n}]\}e^{-r_{z}}\label{1}\\
P_{p_{z}-n,n}(r_{z})&=\{[P^{(z)}_{n}+P^{(y)}_{n}]+[P^{(z)}_{p_{z}-n}+P^{(y)}_{p_{z}-n}]\}e^{-r_{z}}\label{2}\\
Q_{p_{y}-n,n}(r_{y})&=\{[Q^{(z)}_{p_{y}-n}-Q^{(y)}_{p_{y}-n}]-[Q^{(z)}_{n}-Q^{(y)}_{n}]\}e^{-r_{y}}\label{3}\\
P_{p_{y}-n,n}(r_{y})&=\{[P^{(z)}_{p_{y}-n}-P^{(y)}_{p_{y}-n}]+[P^{(z)}_{n}-P^{(y)}_{n}]\}e^{-r_{y}}.\label{4}
\end{align}
can be written in a more compact way using the generalized quadratures $A(\theta)=(ae^{-i\theta}+\ad e^{i\theta})/\sqrt2$:
\begin{align}
A_{p_{z}-n,n}(\theta,r_{z})&=\{[A^{(z)}_{n}(\theta)+A^{(y)}_{n}(\theta)]-[A^{(z)}_{p_{z}-n}(-\theta)+A^{(y)}_{p_{z}-n}(-\theta)]\}e^{-r_{z}} \label{5}\\
A_{p_{y}-n,n}(\theta,r_{y})&=\{[A^{(z)}_{n}(\theta)-A^{(y)}_{n}(\theta)]-[A^{(z)}_{p_{y}-n}(-\theta)-A^{(y)}_{p_{y}-n}(-\theta)]\}e^{-r_{y}}. \label{6}
\end{align}
One can see that $\theta=0$ yields Eqs.~(\ref1) and (\ref3) whereas $\theta=\pi/2$ yields Eqs.~(\ref2) and (\ref4). It is worth noting that the squeezing is independent of $\theta$~\cite{Leuchs2009,Bloomer2010}, hence any value of $\theta$ will do. (However, it is still important to measure at both angles in quadrature, say for the EPR paradox or entanglement in general, since the single-mode $[A(\theta),A(\theta\pm\frac{\pi}2)]\neq0$.)

Look at the Y-pump-centered nullifiers first:
\begin{equation}
  A_{-}(\theta)=[A(\theta)_{n_3 z}-A(-\theta)_{n_4 z}]-[A(\theta)_{n_3 y}- A(-\theta)_{n_4 y}]
  \label{GYS}
\end{equation}
. Write it into the two quadrature nullifier form:
\begin{equation}
Q_{-}(n_3, n_4)=(Q_{n_3 z}-Q_{n_4 z})-(Q_{n_3 y}- Q_{n_4 y})
\end{equation}
\begin{equation}
P_{-}(n_3, n_4)=(P_{n_3 z}+P_{n_4 z})-(P_{n_3 y}+ P_{n_4 y})
\label{Ynullifier}
\end{equation}
where frequency indexes $n_3$ and $n_4$ satisfy the phase matching condition for $yyy$ crystal $n_3 + n_4 = p_{y}$. We checked that both $Q_{-}$ and $P_{-}$ have the same squeezing level by changing the phase of the EOM's driving signal, and this is because the value of phase $\theta$ in Eq.~\ref{GYS} does not change the squeezing.

Similarly, the Z-pump-centered nullifier is:
\begin{equation}
  A_{+}(\theta)=[A(\theta)_{n_1 z}-A(-\theta)_{n_2 z}]+[A(\theta)_{n_1 y}- A(-\theta)_{n_2 y}]
  \label{GZS}
\end{equation}
where $n_1 + n_2 = p_{z}$. Write in the quadrature form:
\begin{equation}
Q_{+}(n_1, n_2)=(Q_{n_1 z}-Q_{n_2 z})+(Q_{n_1 y}- Q_{n_2 y})
\end{equation}
\begin{equation}
P_{+}(n_1, n_2)=(P_{n_1 z}+P_{n_2 z})+(P_{n_1 y}+ P_{n_2 y})
\label{Znullifier}
\end{equation}

Let us look at four modes $n_{3z}$, $n_{4z}$, $n_{3y}$, $n_{4y}$ and their separability conditions.

\subsubsection{One mode- three mode bipartitions}
\paragraph{($n_{3z}$) separable from ($n_{3y}$, $n_{4y}$, $n_{4z}$)}
If mode $n_{3z}$ (the resonant mode with frequency index $3$ and $z$ polarization) is separable from the other three modes, the variances of the nullifiers satisfy the inequality:
\begin{equation}
(\Delta Q_{-}(n_3, n_4))^2+(\Delta P_{-}(n_3, n_4))^2 \geq \frac{1}{2}(|1|+|-1+1-1|)=1
\end{equation}

\paragraph{($n_{4z}$) separable from ($n_{3z}$, $n_{3y}$, $n_{4y}$)}
If mode $n_{4z}$ is separable from the other three modes, the variances of the nullifiers satisfy the inequality:
\begin{equation}
(\Delta Q_{-}(n_3, n_4))^2+(\Delta P_{-}(n_3, n_4))^2 \geq \frac{1}{2}(|-1|+|1+1-1|)=1
\end{equation}

\paragraph{($n_{3y}$) separable from ($n_{3z}$, $n_{4z}$, $n_{4y}$)}
If mode $n_{3y}$ is separable from the other three modes, the variances of the nullifiers satisfy the inequality:
\begin{equation}
(\Delta Q_{-}(n_3, n_4))^2+(\Delta P_{-}(n_3, n_4))^2 \geq \frac{1}{2}(|1|+|1-1-1|)=1
\end{equation}

\paragraph{($n_{4y}$) separable from ($n_{3z}$, $n_{4z}$, $n_{3y}$)}
If mode $n_{4y}$ is separable from the other three modes, the variances of the nullifiers satisfy the inequality:
\begin{equation}
(\Delta Q_{-}(n_3, n_4))^2+(\Delta P_{-}(n_3, n_4))^2 \geq \frac{1}{2}(|-1|+|1-1+1|)=1
\end{equation}

\subsubsection{Two-mode bipartitions}
\paragraph{($n_{3z}$, $n_{3y}$) separable from ($n_{4z}$, $n_{4y}$)}
If modes $n_{3z}$ and $n_{3y}$ are separable from modes $n_{4z}$ and $n_{4y}$, the variances of the nullifiers satisfy the inequality:
\begin{equation}
(\Delta Q_{-}(n_3, n_4))^2+(\Delta P_{-}(n_3, n_4))^2 \geq \frac{1}{2}(|1+1|+|-1-1|)=2
\end{equation}

\paragraph{($n_{3z}$, $n_{4z}$) separable from ($n_{3y}$, $n_{4y}$)}
If modes $n_{3z}$ and $n_{4z}$ are separable from modes $n_{3y}$ and $n_{4y}$, the variances of the nullifiers satisfy the inequality:
\begin{equation}
(\Delta Q_{-}(n_3, n_4))^2+(\Delta P_{+}(n_3, n_5))^2 \geq \frac{1}{2}(|1+0|+|-1+0|)=1
\end{equation}
where $n_3 + n_5 = n_{z pump}$ and $P_{+}(n_3, n_5)$ is a $z$ pump centered nullifier.

\paragraph{($n_{3z}$, $n_{4y}$) separable from ($n_{4z}$, $n_{3y}$)}
If modes $n_{3z}$ and $n_{4y}$ are separable from modes $n_{4z}$ and $n_{3y}$, the variances of the nullifiers satisfy the inequality:
\begin{equation}
(\Delta Q_{-}(n_3, n_4))^2+(\Delta P_{+}(n_3, n_5))^2 \geq \frac{1}{2}(|1+0|+|0-1|)=1
\end{equation}

\subsubsection{Sufficient conditions for inseparability}
The inequalities for each case above are necessary conditions for separability, and a violation of them leads to the sufficient conditions for inseparability. A sufficient condition for the inseparability for all the cases is that the sum of the $P$ and $Q$ nullifiers' variances be smaller than one: $(\Delta Q_{-}(n_3, n_4))^2+(\Delta P_{-}(n_3, n_4))^2 < 1$ and $(\Delta Q_{-}(n_3, n_4))^2+(\Delta P_{+}(n_3, n_5))^2 < 1$. When these sufficient conditions are satisfied, the four modes $n_{3z}$, $n_{4z}$, $n_{3y}$ and $n_{4y}$ are not separable into any subsystems and thus they are entangled. Similar results apply to the Z-pump-centered four modes $n_{1z}$, $n_{2z}$, $n_{1y}$ and $n_{2y}$. And once every four modes are inseparable and their overlapping neighboring four modes are inseparable as well, the whole wire's modes are inseparable because of the transitive property of each 4-mode unit's inseparability. A stronger but simpler sufficient condition for the overall inseparability can be chosen as
\begin{equation}
(\Delta A_{+}(\theta))^2 < \frac{1}{2}
\end{equation}
\begin{equation}
 (\Delta  A_{-}(\theta))^2 < \frac{1}{2}
\end{equation}
This corresponds to the $-3$ dB squeezing level for $A_{+}$ and $A_{-}$, and when the squeezing level is more than this threshold all the modes are inseparable, as we've shown experimentally.

\section{Experimental setup: equipment and procedures}

This completes the description of the experimental setup in the main text and details some of the experimental procedures that were used.

\subsection{Equipment}

The PPKTP crystals were provided by Raicol, Inc., and were 10 mm-long, $x$-cut, periodically poled at 9 $\mu$m so as to quasiphasematch~\cite{Fejer1992,Pooser2005,Pysher2010}
$zzz$ PDC. They were antireflection coated by Advanced Thin Films at 1064 nm (for both polarizations) and 532nm, and mounted oriented at $90^{\circ}$ from each other in the ($yz$) plane. Each crystal was temperature-controlled to a few tenths of a millidegree by using Wavelength Electronics servo loop chips, and the temperature was tuned within the phase matching bandwidth so as to equate the optical paths at each polarization.

The OPO mirrors were fabricated by Advanced Thin Films. The cavity was formed by two concave mirrors (50 mm radius) and two flat mirrors, one of which the output coupler of transmissivities of 5\% at 1064 nm and 0.05\% at 532 nm. All other mirrors have transmissivities of near-zero at 1064 nm and near-unity at 532 nm. The OPO cavity length was actively stabilized by locking it to a weak counterpropagating LO beam via a Pound-Drever-Hall servo loop. Our servo loops were all built in house, except for one Vescent D2-125 module that was occasionally used. The bowtie resonator had two beam waists, of 31 $\mu$m (between the curved mirrors) and 131 $\mu$m (between the flat mirrors), where the two PPKTP nonlinear crystals were placed.

Great care was taken to verify that there is no polarization crosstalk between the two crystals by generating the second harmonic of a 1064 nm seed laser beam modematched to the OPO cavity, and by checking the absence of $y$($z$)-polarized radiation at 532 nm in the presence $z$($y$)-polarized seed at 1064nm.

We also noticed that the ring OPO, when seeded by a laser beam, exhibited a retroreflected beam from a cavity mode counterpropagating to the seed mode, and whose power can reach the order of 10\% of the incident seed's power. We found that this counterpropagating mode stemmed from residual reflections on the crystal faces, which created an intra-OPO system of coupled cavities. We managed to minimize this resonant retroreflection from the OPO by slightly angling the crystals in the OPO cavity.

Two frequency-doubled, ultrastable continuous-wave Nd:YAG lasers (Innolight ``Diabolo'' 1W), of frequency linewidth 1 kHz at 532 nm, were used for the pump fields. The lasers were phaselocked together at a frequency difference $2m\Delta\omega$, with $m=1,2$. This was achieved by a standard superheterodyne setup: one of the lasers was controlled via its laser crystal piezotransducer so as to phaselock the lasers' beat note to the stable radiofrequency delivered by an Agilent E8247C CW signal generator. The two pump beams then entered the OPO through different paths to access the $yyy$ and $zzz$ PPKTP crystals separately.

The two-tone balanced homodyne detection system used $95\%$-efficient  JDSU ETX500T InGaAs photodiodes. Another Nd:YAG continuous-wave laser (JDSU Lightwave Electronics Model 126) provided the LO and the OPO locking beam. The LO frequencies were generated by phase EOM sidebands from a Photline NIR-MPX modulator, driven by a Hittite HMC-T2100 generator. The filter cavity was locked such that the first order harmonics of the LO sideband will transmit to beat with the particular frequencies we intend to measure, and the zero and second order will reflect.

The LO laser was phaselocked to one of the fundamental pump lasers by shifting the LO frequency by 70 MHz with an IntraAction ATM-801A2 acousto-optic modulator and locking the resulting beat note to a Hewlett Packard 8648A signal generator, phaselocked to another, identical, signal generator which was the 70 MHz source. That way, having both generators at 70 MHz ensured both lasers could be locked at the same frequency. When we needed a frequency offset to check the graph, we simply shifted the frequency of the first generator.

The squeezing measurements were performed on an Agilent E4402B spectrum analyzer, the detection network being made of Mini-Circuits components.

\subsection{Electronic noise correction}\label{noisecorrection}
The squeezing traces we show in the figures are the original raw measurements without any correction. The actual squeezing should be more after taking account the effects of the detector's electronic noise. Electronic noise, also known as dark noise, is about $13$ dB below the shot noise (shown in Figure \ref{darknoise}).
\begin{figure}
\centerline{\includegraphics[width=0.45\columnwidth]{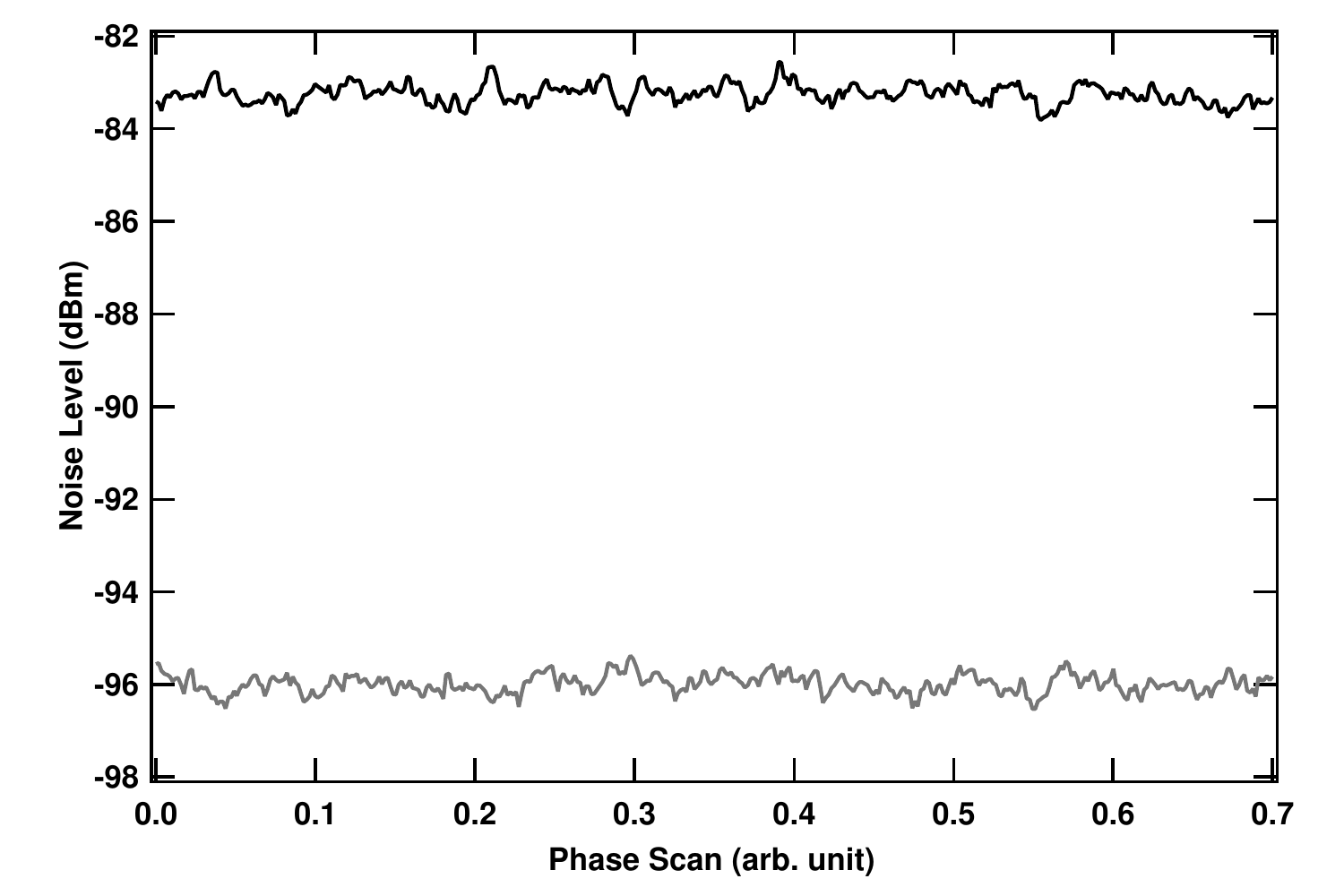}}
\vglue -0.1in
\caption{Electronic noise measurement. Black (top trace): LO shot noise; grey (bottom trace): electronic noise. }
\vglue -0.1in
\label{darknoise}
\end{figure}
The actual squeezing level is
\begin{equation}
S_{act}=10\log{\eta_{act}}=10\log{\frac{V_{sq}}{V_{sn}}}
\end{equation}
While the experimentally measured squeezing level, contaminated by the electronic noise, is
\begin{equation}
S_{exp}=10\log{\eta_{exp}}=10\log{\frac{V_{sq}+V_{en}}{V_{sn}+V_{en}}}=10\log{\frac{\eta_{act}+\frac{V_{en}}{V_{sn}}}{1+\frac{V_{en}}{V_{sn}}}}
\end{equation}
where $V_{sq}$ is the variance of the squeezing signal, $V_{sn}$ is the variance of the shot noise and $V_{en}$ is the variance of the electronic noise.
So we have
\begin{equation}
\eta_{act}=(\eta_{exp}-1)\frac{V_{en}}{V_{sn}}+\eta_{exp}
\end{equation}
Given our experiment's squeezing level, after the correction, the squeezing level increases $>0.2$ dB, as shown in Table 1.
\begin{figure}
\centerline{\includegraphics[width=0.6\columnwidth]{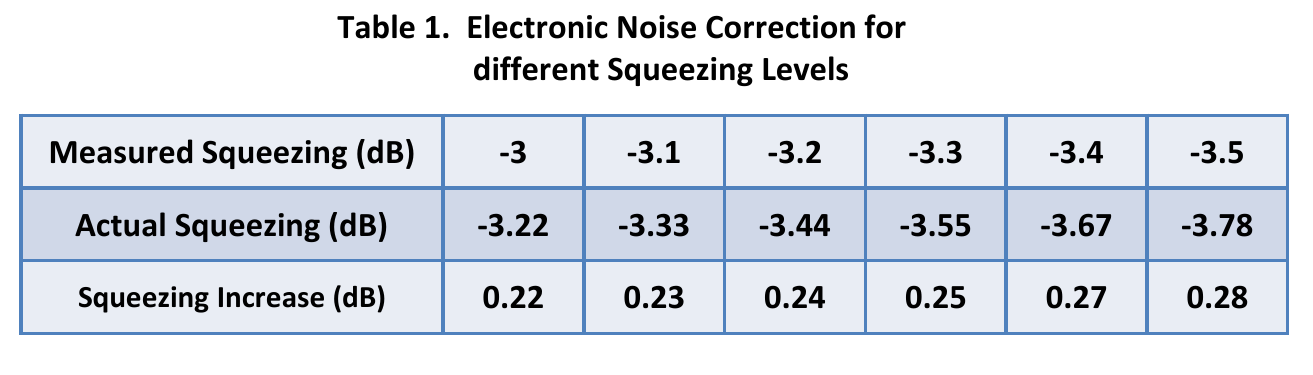}}
\vglue -0.1in
%\caption{Table}
\vglue -0.1in
\label{table}
\end{figure}

\subsection{Role of the EOM phase in nullifier selection}

As was already mentioned, the nullifiers can be written in a more compact way using the generalized quadratures, Eqs.~{\ref5) \& (\ref6). Now tracking the experimental phases, i.e., the LO phase $\theta_{LO}$ and the EOM phase $\theta_{o}$ (see Fig.~4 of main text), it can be shown (see also supplemental material of~\cite{Pysher2011s}) that squeezing in Eqs.~(\ref1,\ref2) is always obtained for values of $\theta_{LO}$ which are multiples of $\pi$. At such values, the operators are then given by
 \begin{align}
A_{p_{z}-n,n}(\theta,r_{z})&=\{[A^{(z)}_{n}(\theta_{o})+A^{(y)}_{n}(\theta_{o})]-[A^{(z)}_{p_{z}-n}(-\theta_{o})+A^{(y)}_{p_{z}-n}(-\theta_{o})]\}e^{-r_{z}} \\
A_{p_{y}-n,n}(\theta,r_{y})&=\{[A^{(z)}_{n}(\theta_{o})-A^{(y)}_{n}(\theta_{o})]-[A^{(z)}_{p_{y}-n}(-\theta_{o})-A^{(y)}_{p_{y}-n}(-\theta_{o})]\}e^{-r_{y}}
\end{align}
and then changing $\theta_{o}$ by $\pi/2$ provides a way to measure, say, Eqs.~(\ref1,\ref2). In practice, we tuned $\theta_{o}$ by swapping coaxial cables differing in length by 1 foot, between the EOM and its driver. Indeed, since the EOM frequency is always $(n+\frac12)\Delta\omega\simeq(n+\frac12)$ GHz, the phase shift is $(n+\frac12)10\pi \ell$ in an RG-58 cable of length $\ell$, with a $2c/3$ propagation velocity in the cable. Choosing $\ell=0.3$ m, or 1 foot, therefore yields close to the desired $\pm\pi/2$ phase shift when $n=0$. As shown in Fig.~\ref{fig:beltphase}, the squeezing performance is the same for both nullifiers, as expected. At other frequencies, a 1 foot cable won't exactly yield a quadrature phase shift, yet the squeezing abides (Fig.~\ref{fig:beltphase2}).

%\newpage

\begin{figure}
\centerline{\includegraphics[width=0.85\columnwidth]{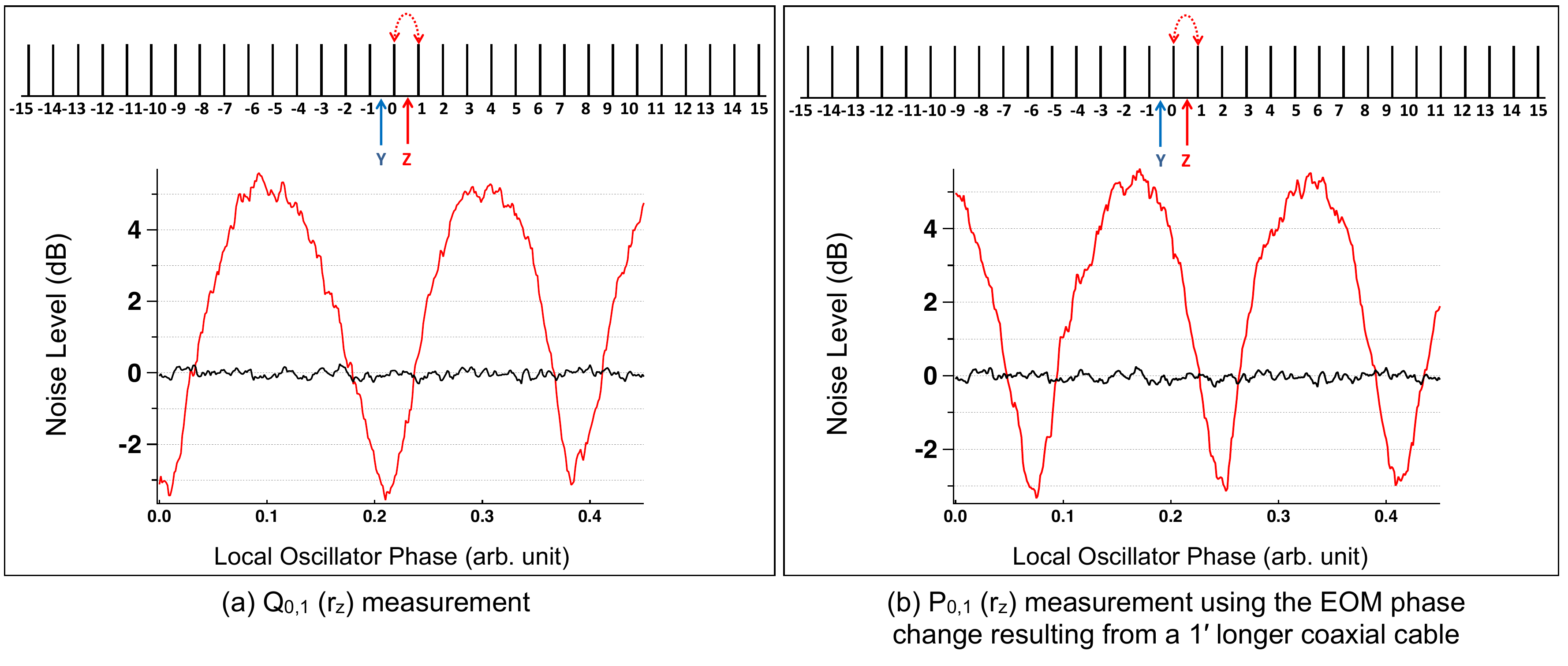}}
\vglue -0.1in
\caption{Squeezing measurements for the same generalized quadrature nullifier at phases in quadrature. The squeezing performances are the same in both cases. The horizontal axis is the scan of the local oscillator's phase in arbitrary units. }
\vglue -0.1in
\label{fig:beltphase}
\end{figure}
\begin{figure}
\centerline{\includegraphics[width=0.9\columnwidth]{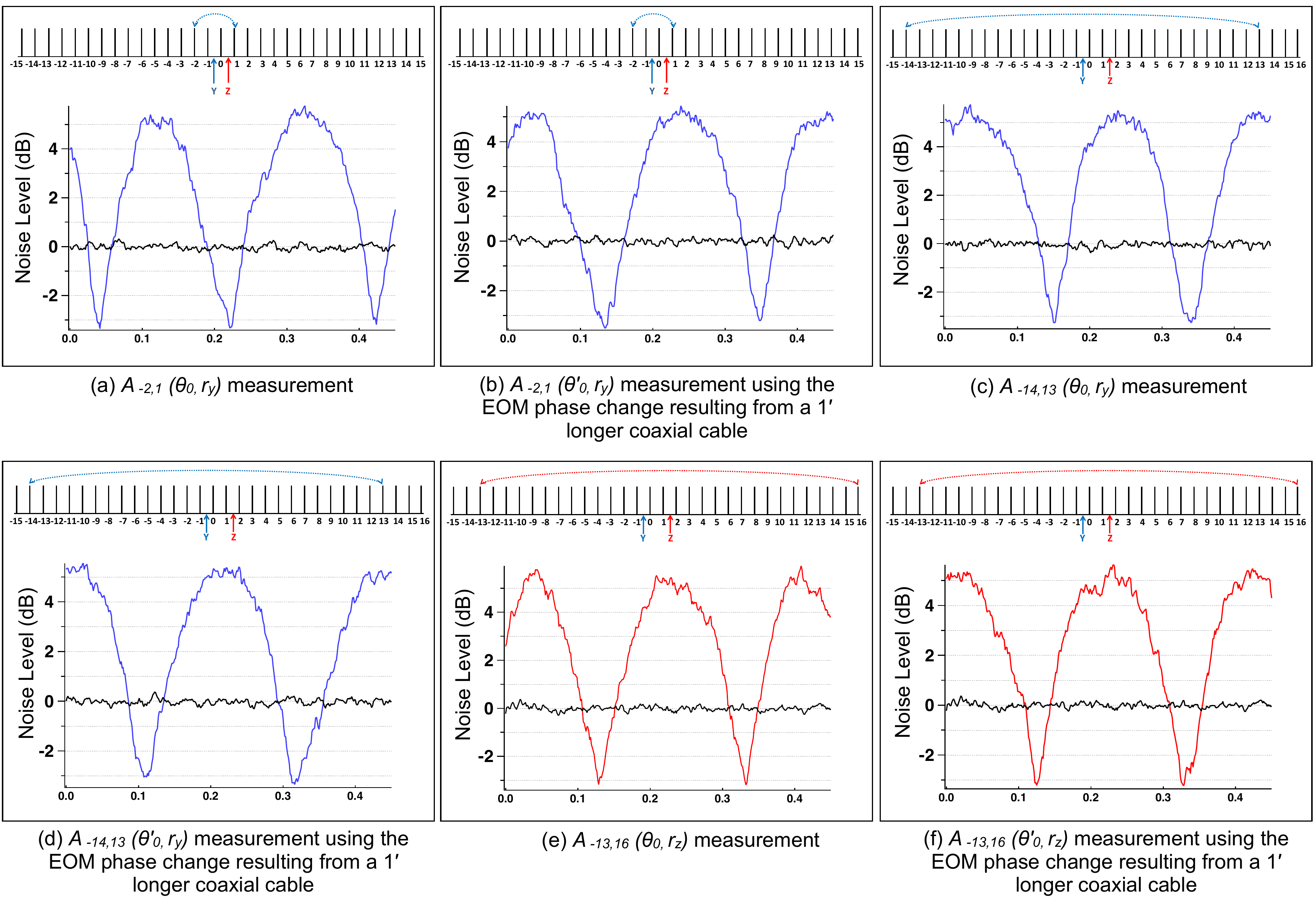}}
\vglue -0.1in
\caption{Squeezing measurements for the same generalized quadrature nullifier at different phases $\theta_{o},\theta'_{o}$, separated by 1 foot length difference of coaxial cable. The squeezing performances are the same in all cases. The horizontal axis is the scan of the local oscillator's phase in arbitrary units. }
\vglue -0.1in
\label{fig:beltphase2}
\end{figure}

%\newpage

\section{Complete measurement data}\label{squeezing}
We have measured and confirmed the dual-rail wire structure up to 60 modes for one-wire case and 30 modes for each wire for two-wire case. We change the local oscillator's frequency combined with its sidebands modulated by EOM to precisely pinpoint which modes we are measuring. Due to the limited space in the main  text we will show all of the measuring results here (all traces are the raw measurements).

\subsection{One-wire case}
\subsubsection{Squeezing measurements for the one-wire case}
Start from the squeezing measurement for the one-wire case. The y pump centered 4-mode nullifier (Eq.~\ref{GYS}) measurements are shown in Figure~\ref{fig:OnebeltY}, and the z pump centered 4-mode nullifier (Eq.~\ref{GZS}) measurements are shown in Figure~\ref{fig:OnebeltZ}. These are all original measurements without any noise correction; after the homodyne detector's electronic noise correction the squeezing will increase 0.2 dB more for all the cases. The total number of modes we measured in one wire is 60, which is only limited by the EOM measurement ability (while the actual number of modes in our wire should be many orders of magnitude more).

\begin{figure}[t]
\centerline{\includegraphics[width=0.9\columnwidth]{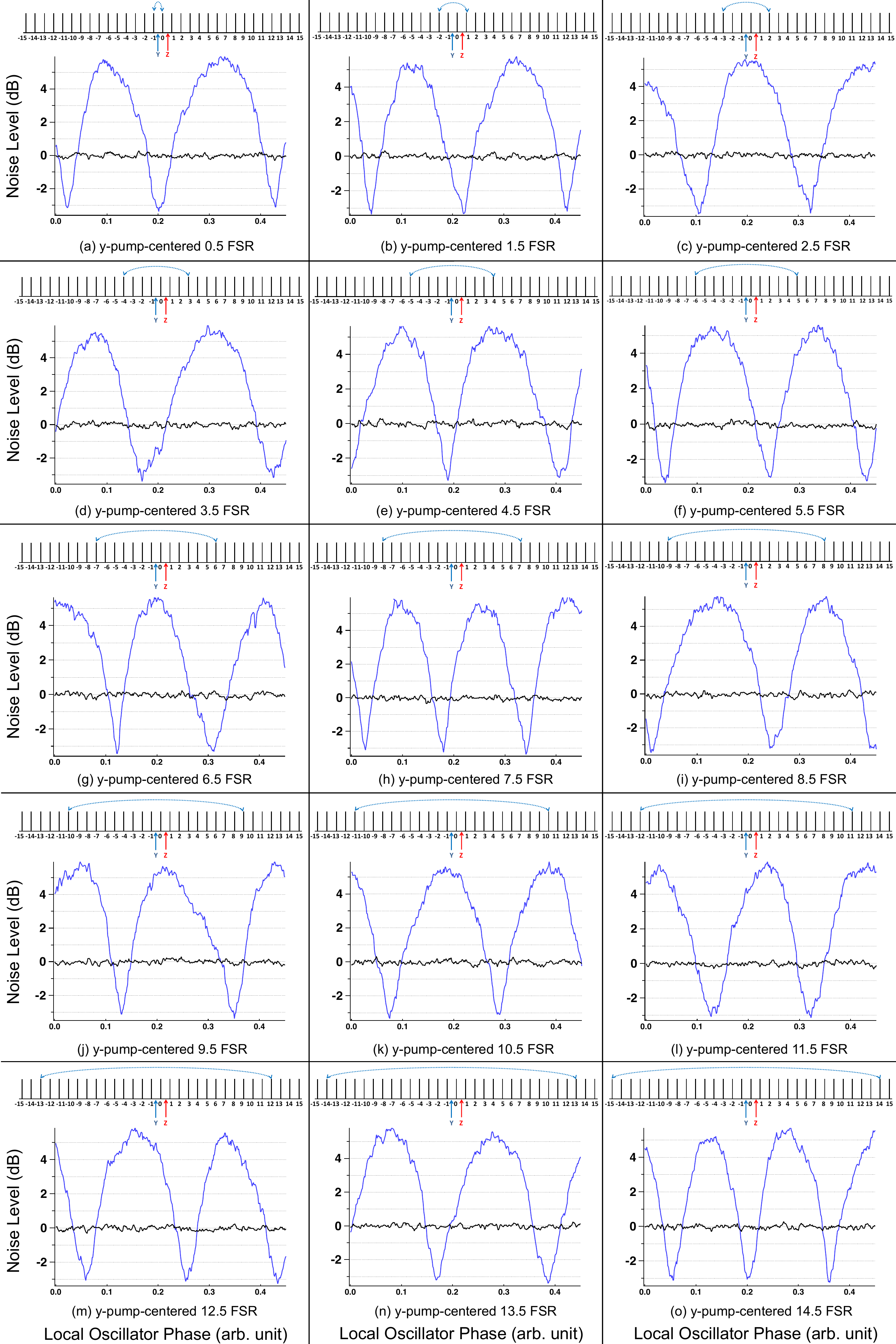}}
\vglue -0.1in
\caption{One-wire case y pump centered nullifiers squeezing measurements.}
\vglue -0.1in
\label{fig:OnebeltY}
\end{figure}

\begin{figure}[t]
\centerline{\includegraphics[width=0.9\columnwidth]{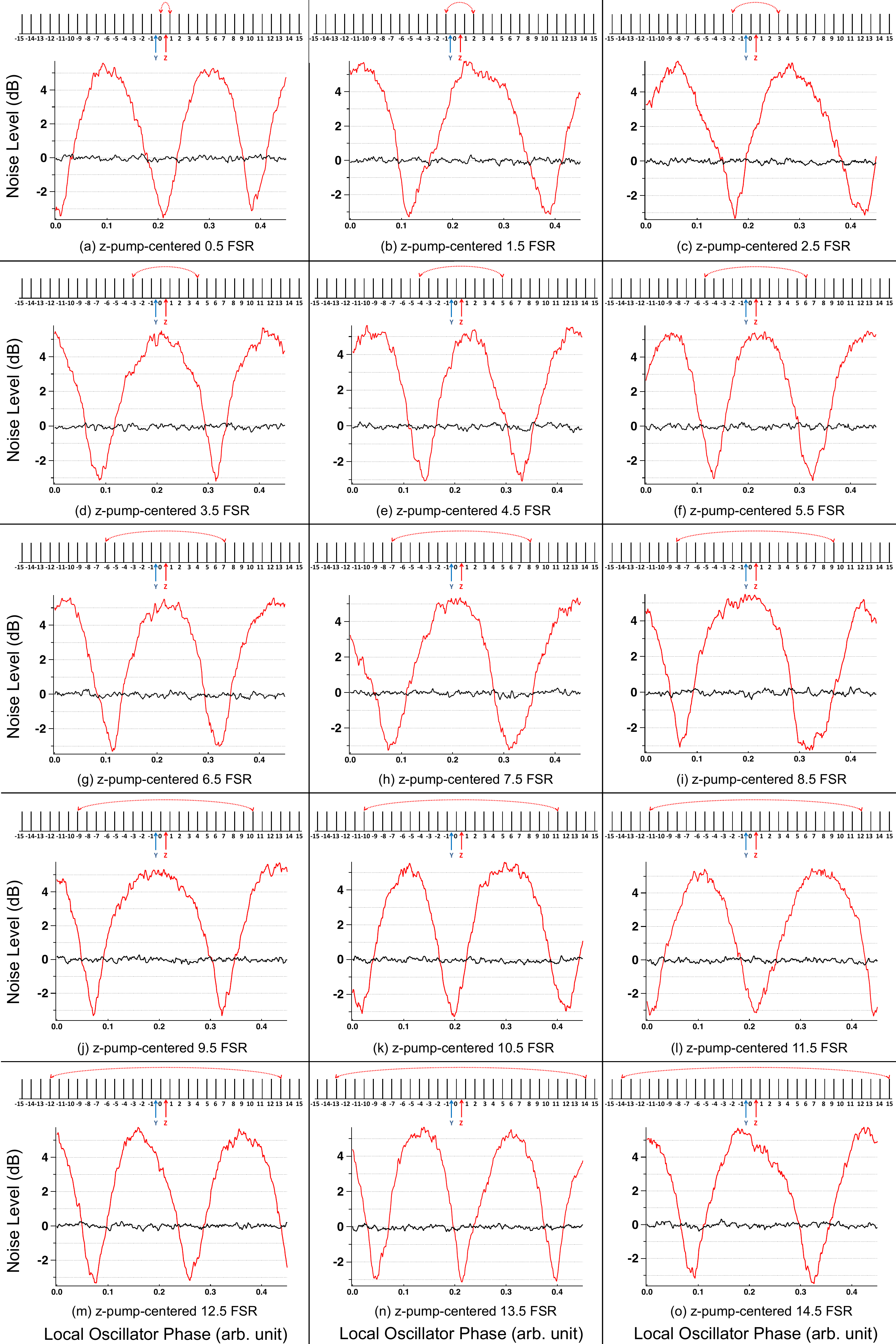}}
\vglue -0.1in
\caption{One-wire case z pump centered nullifiers squeezing measurements.}
\vglue -0.1in
\label{fig:OnebeltZ}
\end{figure}

\subsubsection{Wrong-frequency checks for the one-wire case}
We set the local oscillator frequency to measure the modes that are not supposed to have connections for some wrong-frequency checks, and it shows that when we are not measuring the right modes (intentionally tuning the local oscillator frequency to other non-pump-symmetric modes), although we are measuring the same nullifiers form, we do not obtain squeezing. We calculated that the variances of the nullifiers at wrong frequencies are
\begin{align}
(\Delta A_{-}(\theta)_{n_i n_{ii}})^2&=\cosh{2r_y}\\
(\Delta A_{+}(\theta)_{n_i n_{ii}})^2&=\cosh{2r_z},
\end{align}
 where frequency indexes $n_i$ and $n_{ii}$ do not satisfy either of the phase matching condition, so $n_i + n_{ii} \neq p_{Y pump}$ and $n_i + n_{ii} \neq p_{Z pump}$, which means the two frequency indexes $n_i$ and $n_{ii}$ are not symmetric about either pump. It shows that there is only antisqueezing, and the antisqueezing levels are independent of the local oscillator's phase, which agrees with our results, shown in Figure~\ref{fig:Onebeltcheck}.
\begin{figure}[!htb]
\centering
\subfigure[ One-wire Modes (-1) and (-2)]{
\includegraphics[width=0.4\textwidth]{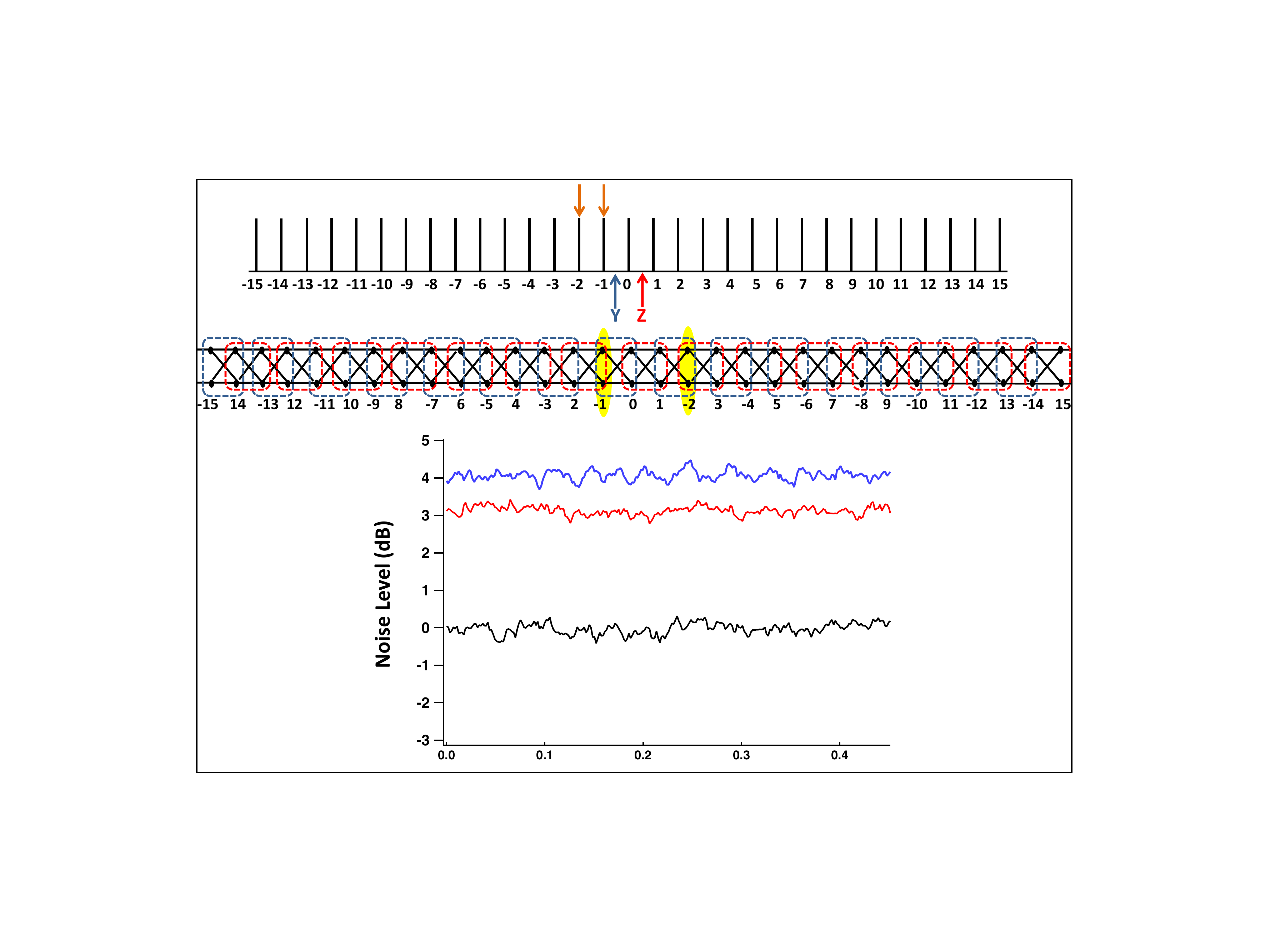}
\label{fig:Onebeltf-1f-2}
  }
\subfigure[ One-wire Modes (-12) and 15]{
\includegraphics[width=0.4\textwidth]{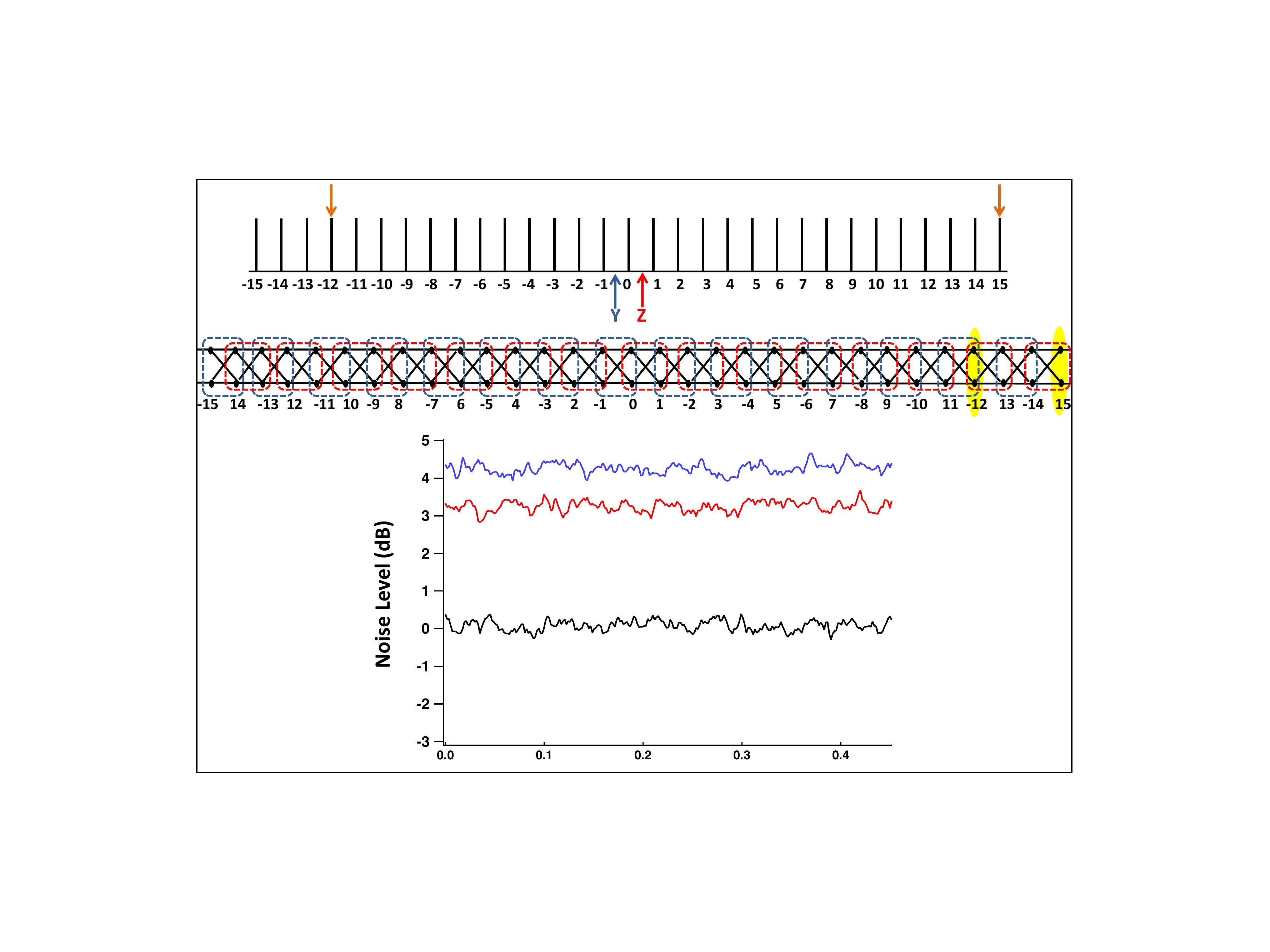}
\label{fig:Onebeltf15f-12}
  }
\subfigure[ One-wire Modes 1 and 2]{
\includegraphics[width=0.4\textwidth]{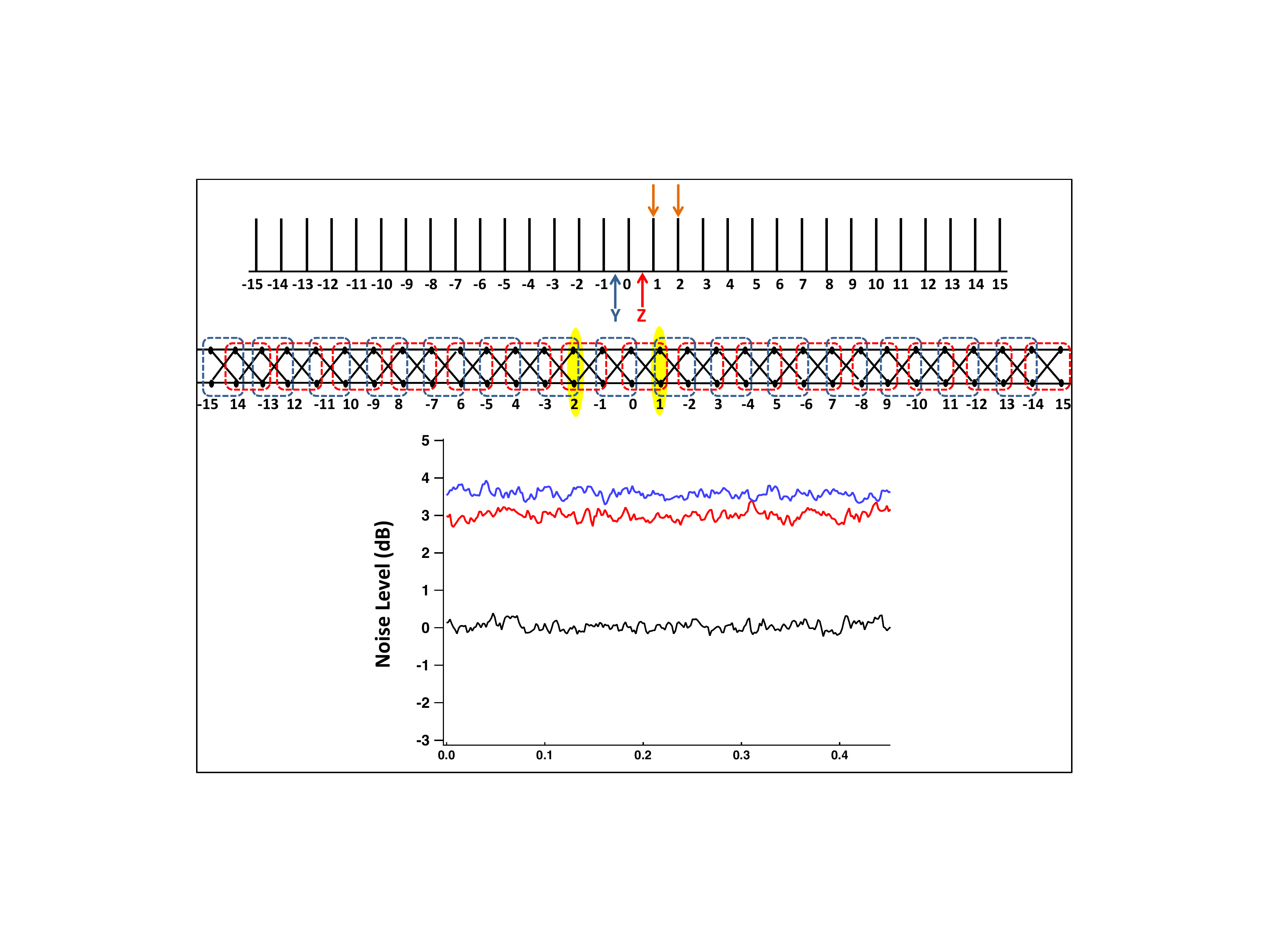}
\label{fig:Onebeltf1f2}
  }
\subfigure[ One-wire Modes vacuum]{
\includegraphics[width=0.4\textwidth]{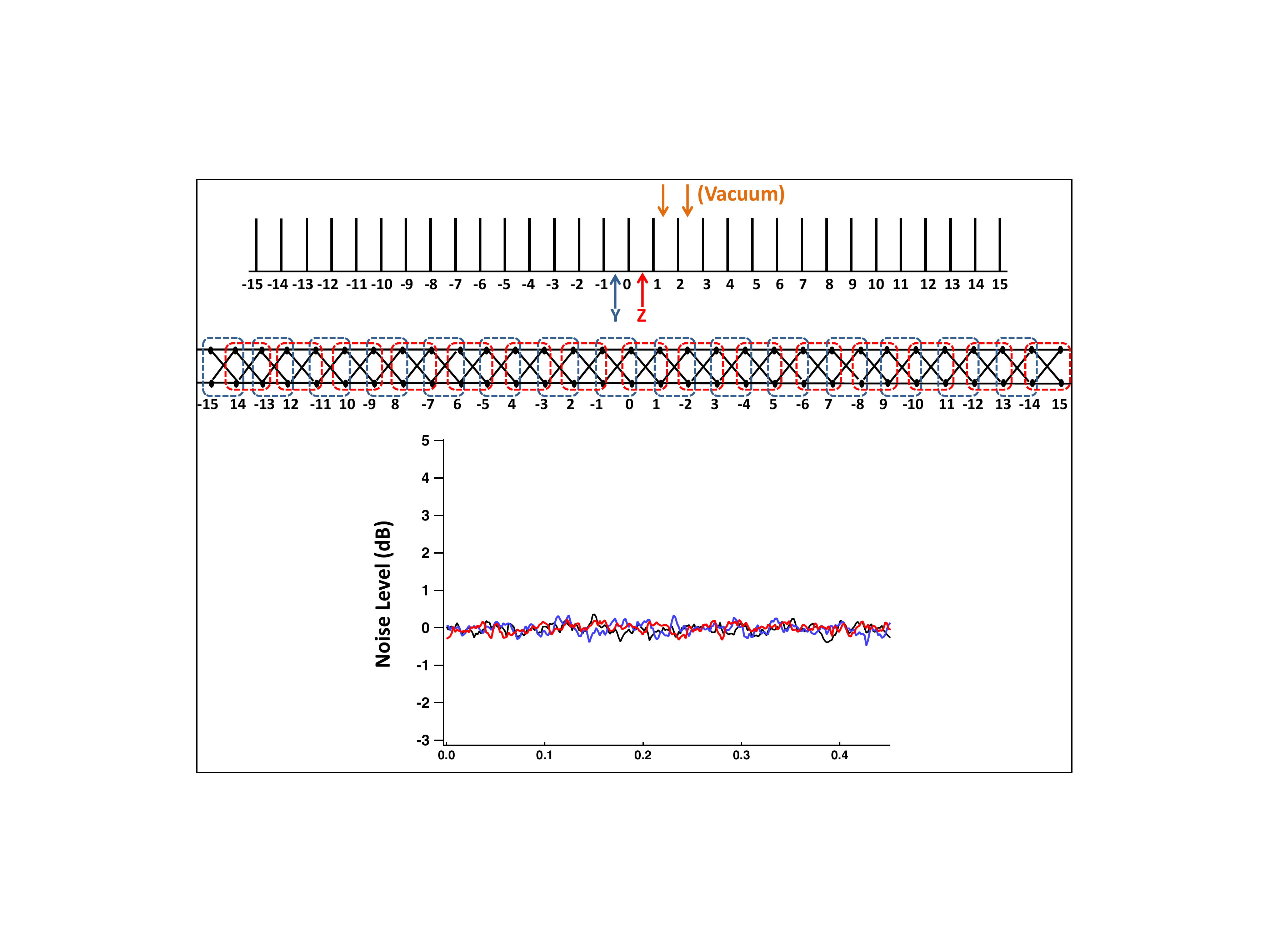}
\label{fig:Onebeltfvac}
  }
\caption{Wrong-frequency measurements for one-wire case.
Antisqueezing was observed between non-neighboring modes in one wire(Fig \ref{fig:Onebeltf-1f-2}, \ref{fig:Onebeltf15f-12}, \ref{fig:Onebeltf1f2}). Shot noise level was measured when the local oscillator frequencies were tuned to that between modes as in Fig.~\ref{fig:Onebeltfvac}. Note that in Fig.~\ref{fig:Onebeltf1f2}, Modes 1 and 2 would have been squeezed in the two-wire case but have no connection here, showing the one-wire and two-wires cases are indeed different from each other. The yellow ellipses indicate the modes we are measuring for each case. Black traces are the shot noise level, blue traces are the y pump centered nullifiers (Eq~\ref{GYS}) and red traces are the z pump centered nullifiers (Eq~\ref{GZS}). In the figure the horizontal axis is the scan of the local oscillator's phase in arbitrary units.}
\vglue -0.1in
\label{fig:Onebeltcheck}
\end{figure}

\subsection{Two-wire case}
Similar to the one-wire case, we also measured the nullifiers whose modes are symmetric about the y- or z- pump respectively, and we also performed the wrong-frequency checks as in the previous case.

\subsubsection{Squeezing measurements for the two-wire case}
We measured the nullifiers whose modes are symmetric about the y- or z- pump respectively and obtained squeezing for all of them. The squeezing levels are constant throughout all the nullifiers indicating no sign of entanglement loss as we move further from the center. Again, the number of modes we measured for each wire in the two-wire case was only limited due to the EOM measurement ability, not the state itself (which, we believe, has many orders of magnitude more modes).

\begin{figure}
\centerline{\includegraphics[width=0.9\columnwidth]{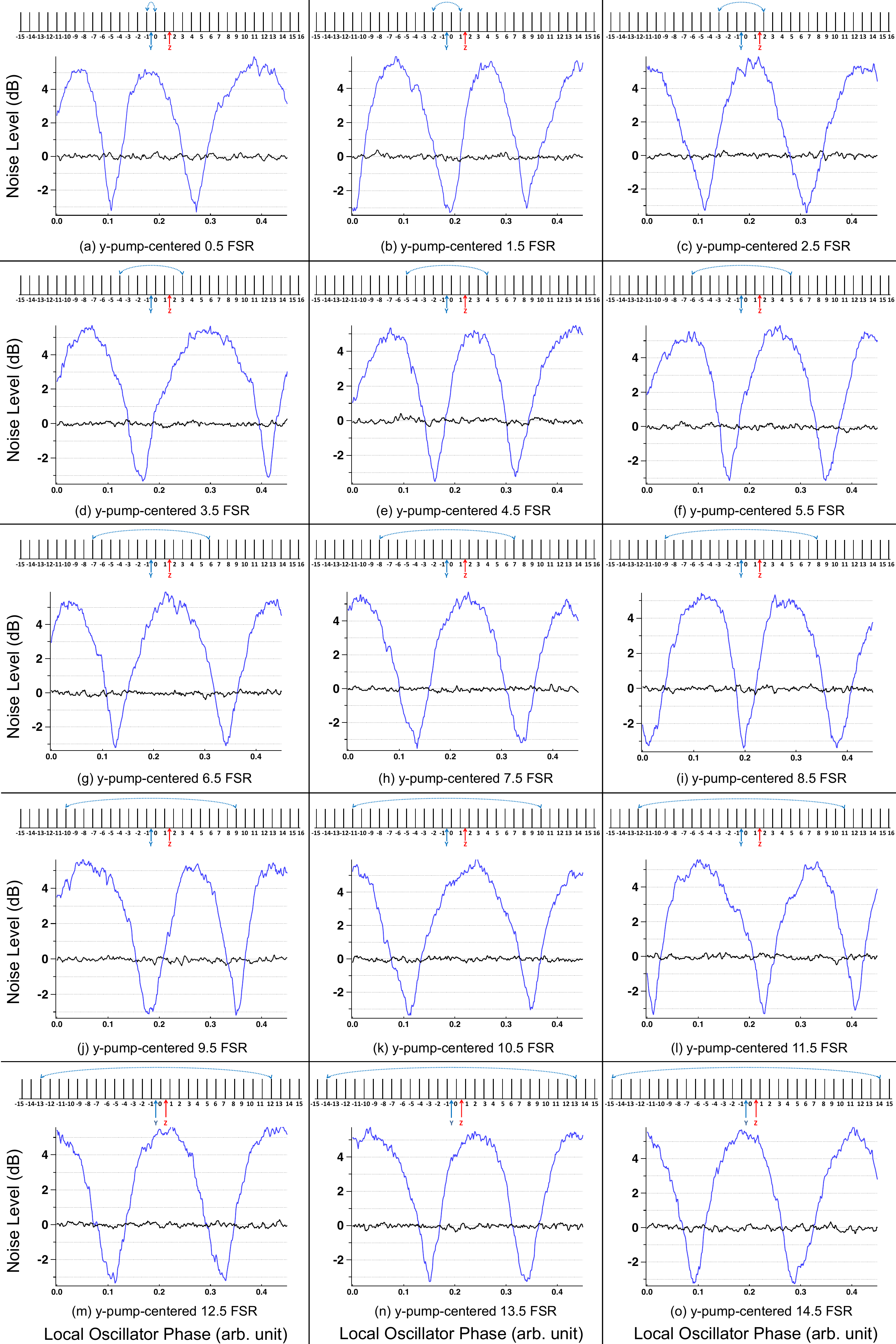}}
\vglue -0.1in
\caption{Two-wire case y pump centered nullifiers squeezing measurements.}
\vglue -0.1in
\label{fig:TwobeltZ}
\end{figure}

\begin{figure}
\centerline{\includegraphics[width=0.9\columnwidth]{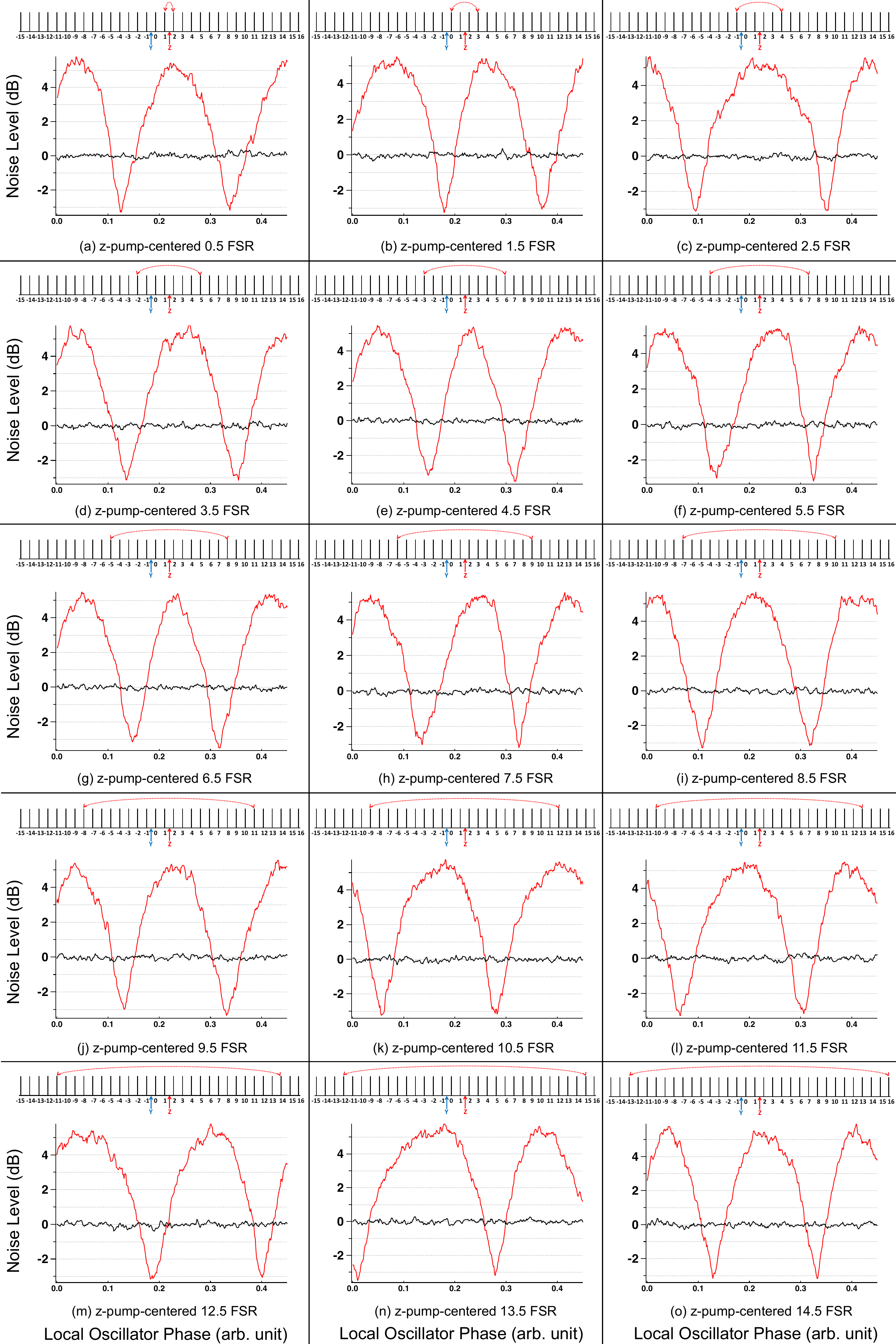}}
\vglue -0.1in
\caption{Two-wire case z pump centered nullifiers squeezing measurements.}
\vglue -0.1in
\label{fig:TwobeltZ}
\end{figure}

\subsubsection{Wrong-frequency checks for the two-wire case}
Similar to the one-wire case, the wrong-frequency checks for the two-wire case only have antisqueezing or shot noise throughout the checks and no squeezing was detected at any time. The wrong-frequency checks show that there is no connection between the two wires and thus they are two independent wires. (Strictly, in order to show the independency of each wire, measurements between all the modes between the two wires are needed, but given the extremely large number of modes in a wire such measurements are tedious and beyond the EOM measure limit so we measured a few to show no sign of connection, which can be generalized.)

\begin{figure}[!htb]
\centering
\subfigure[ Two-wire Modes 1 and 0]{
\includegraphics[width=0.4\textwidth]{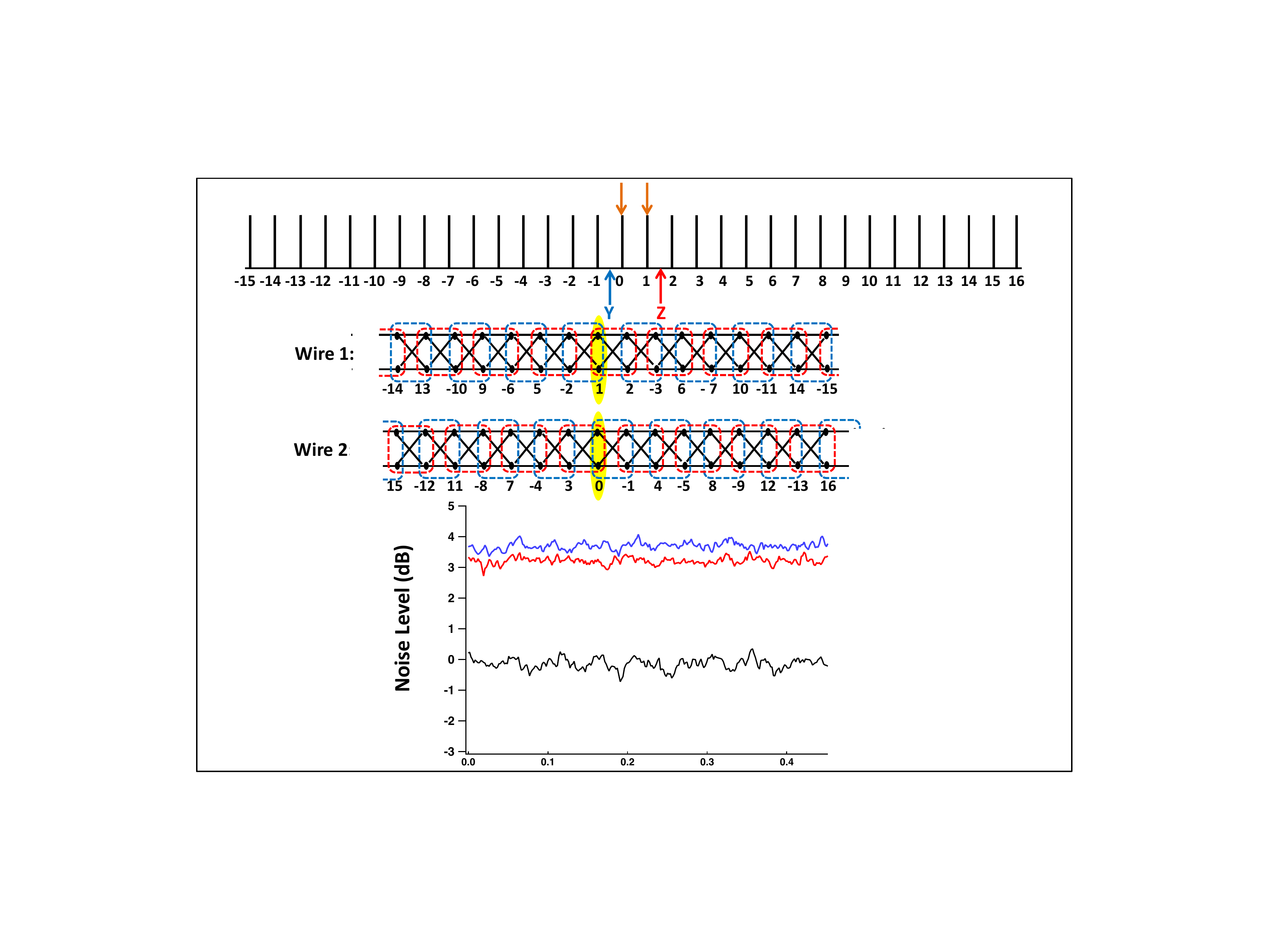}
\label{fig:twobeltsf1f0}
  }
\subfigure[ Two-wire Modes 14 and (-13)]{
\includegraphics[width=0.4\textwidth]{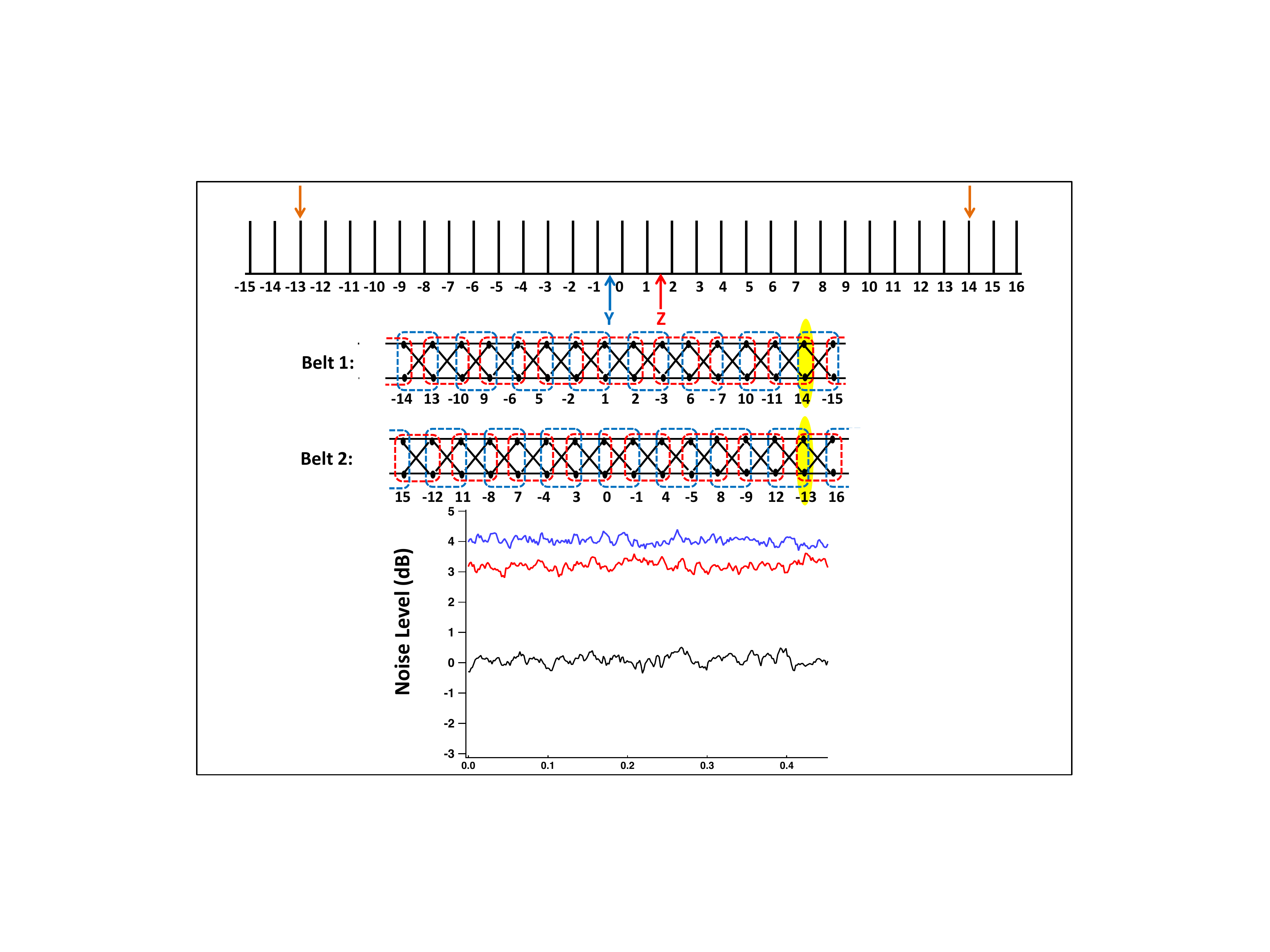}
\label{fig:Twobeltsf14f-13}
  }
\subfigure[ Two-wire Modes (-2) and (-3)]{
\includegraphics[width=0.4\textwidth]{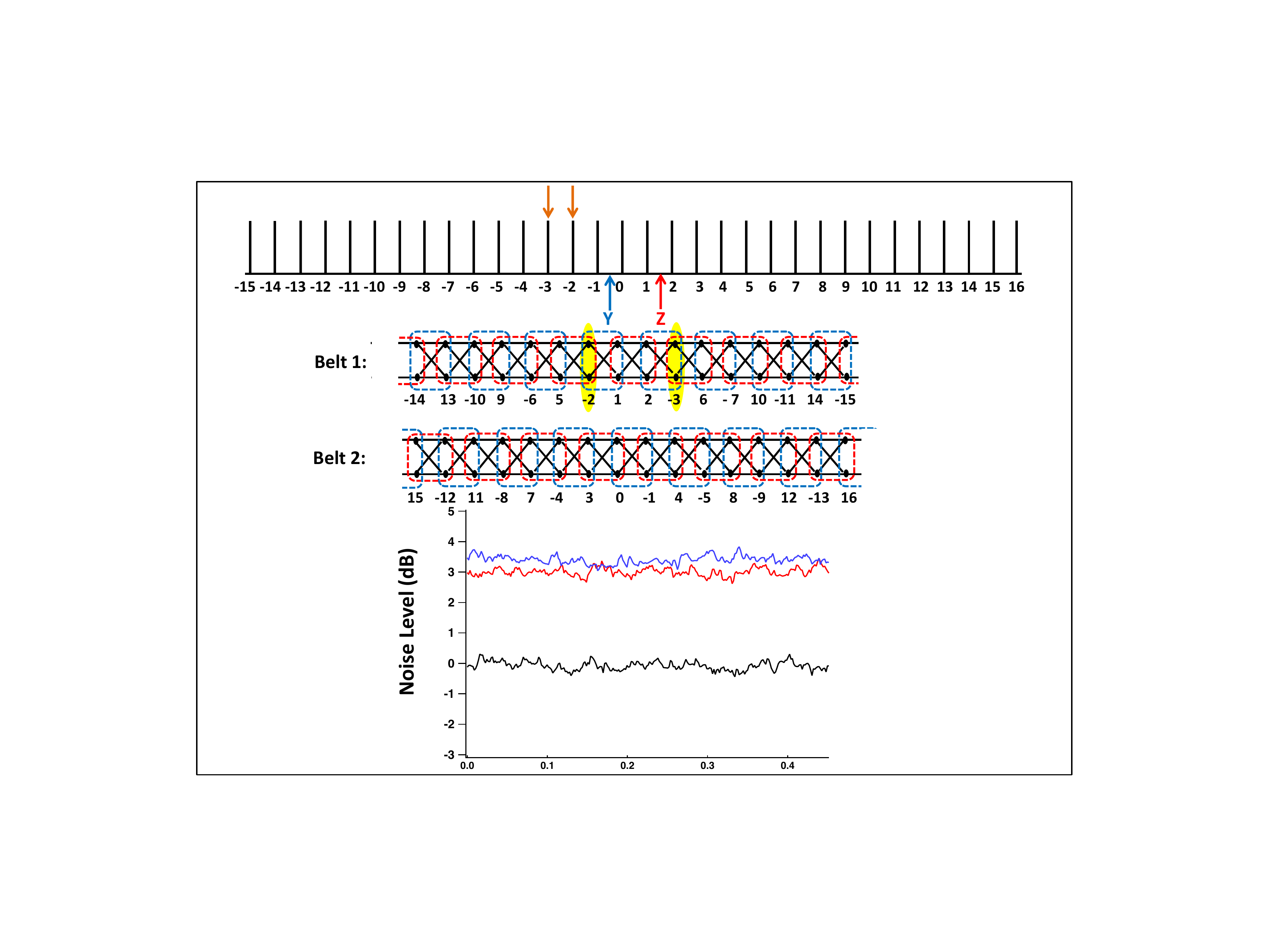}
\label{fig:twobeltsf-2f-3}
  }
\subfigure[ Two-wire Modes vacuum]{
\includegraphics[width=0.4\textwidth]{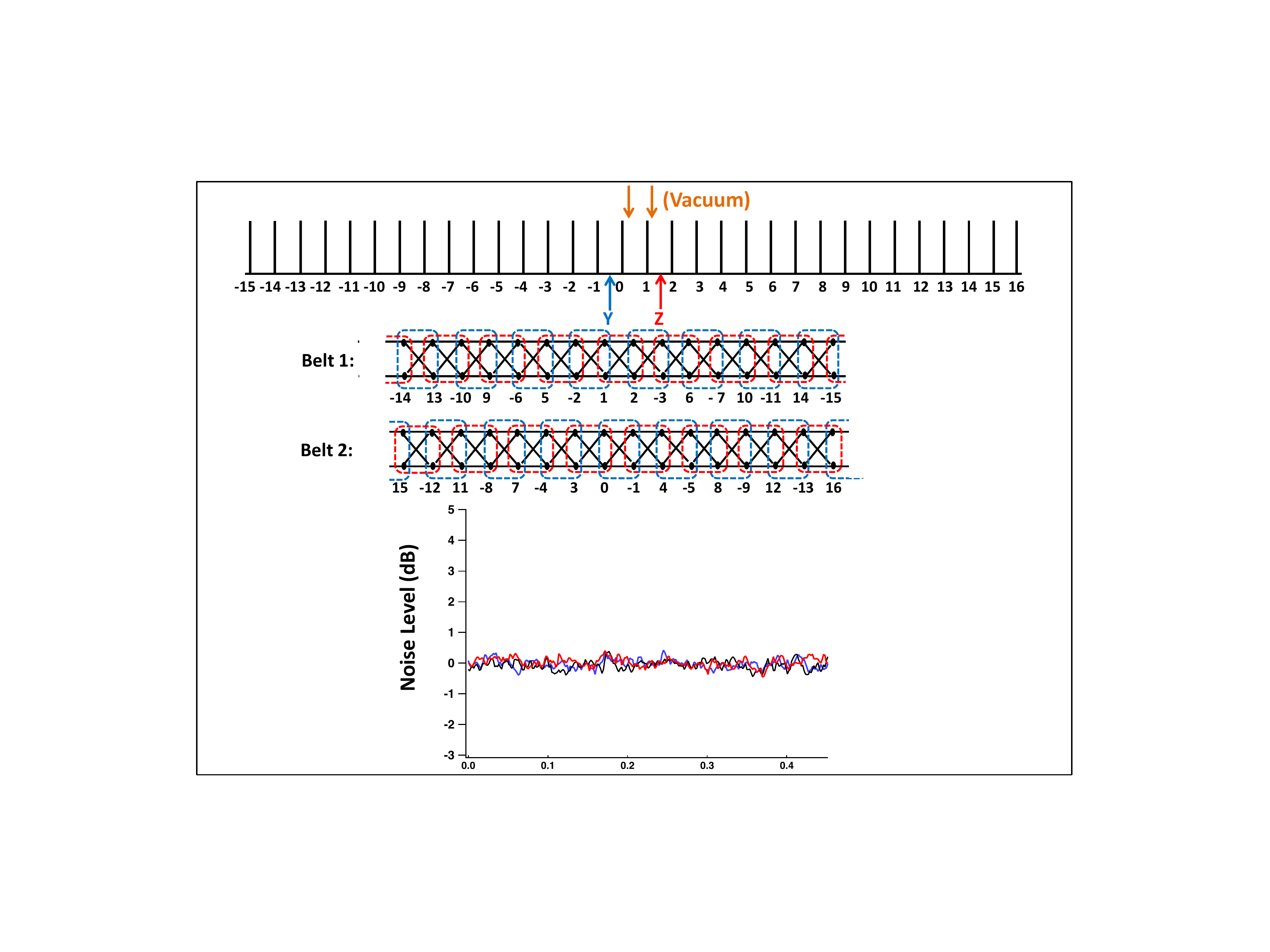}
\label{fig:twobeltsfvac}
  }
\caption{Wrong-frequency measurements for two-wire case.
Antisqueezing was observed either between modes in the same wire (Fig~\ref{fig:twobeltsf-2f-3} or modes in different wires (Fig~\ref{fig:twobeltsf1f0},~\ref{fig:Twobeltsf14f-13}), showing that only the neighboring modes in the same wire are connected, confirming the independent two-wire structure. Shot noise level was obtained when measuring frequencies between modes (Fig~\ref{fig:twobeltsfvac}). Note that in Fig~\ref{fig:twobeltsf1f0} and ~\ref{fig:Twobeltsf14f-13}, Modes 1 and 0 or Modes 14 and (-13) would have been squeezed in the one-wire case but have no connection here. The yellow ellipses indicate the modes we are measuring for each case. Black traces are the shot noise level, blue traces are the y pump centered nullifiers (Eq~\ref{GYS}) and red traces are the z pump centered nullifiers (Eq~\ref{GZS}). In the figure the horizontal axis is the scan of the local oscillator's phase in arbitrary units.}
\vglue -0.1in
\label{fig:twobeltscheck}
\end{figure}

\section{Squeezing imperfections}

Here we investigate the consequences of $r_{z}\neq r_{y}$ to first order. We have the initial nullifiers
\begin{align}
[(Q^{z}_{0}+Q^{y}_{0})&-(Q^{z}_{1}+Q^{y}_{1})]e^{-r_{z}}\label{eq:nbs1}\\
[(P^{z}_{0}+P^{y}_{0})&+(P^{z}_{1}+P^{y}_{1})]e^{-r_{z}}\label{eq:nbs2}\\
[(Q^{z}_{-1}-Q^{y}_{-1})&-(Q^{z}_{0}-Q^{y}_{0})]e^{-r_{y}}\label{eq:nbs3}\\
[(P^{z}_{-1}-P^{y}_{-1})&+(P^{z}_{0}-P^{y}_{0})]e^{-r_{y}}.\label{eq:nbs4}
\end{align}
Assuming $r_{z,y}=r\pm\varepsilon$, $\varepsilon\ll r$, taking the sum and difference of \eq{nbs1} and \eq{nbs3}, and applying a Fourier transform, a.k.a.\ a local $\frac\pi2$ optical phase shift, to mode 0 yields, to first order in $\varepsilon$
\begin{align}
\left\{P^{z}_{0} -\varepsilon P^{y}_{0} - \frac12 [(1-\varepsilon)(Q^{y}_{1}+Q^{z}_{1})+(1+\varepsilon)(Q^{z}_{-1}-Q^{y}_{-1})]\right\}e^{-r}\label{eq:nps1f}\\
\left\{P^{y}_{0} -\varepsilon P^{z}_{0}- \frac12 [(1-\varepsilon)(Q^{y}_{1}+Q^{z}_{1})-(1+\varepsilon)(Q^{z}_{-1}+Q^{y}_{-1})]\right\}e^{-r}\label{eq:nps2f}
\end{align}
So the effect of unbalanced squeezing is a spurious correlation between the $0z$ and $0y$ modes, as well as edge weights of unequal magnitude in the rest of the graph. While these effects can be made arbitrarily small in our experiment by tuning the relative pump intensities, they ought to be kept in mind when evaluating the performance of future quantum processing applications.

\end{document}